\newcommand{\tess}{\textit{TESS}}
\newcommand{\starname}{IRAS 04125+2902}
\newcommand{\planetname}{IRAS 04125+2902\,b}
\newcommand{\companionname}{TIC 56658273}

\newcommand{\fbol}{$F_{\mathrm{bol}}$}

\newcommand{\um}{$\mu$m}
\newcommand{\gaia}{\textit{Gaia}}
\newcommand{\prot}{$P_{\rm{rot}}$}

\documentclass{nature3}
\usepackage{graphicx}
\usepackage{float}
\usepackage{verbatim}
\usepackage{hyperref}
\usepackage{amsmath}
\usepackage{amssymb}
\usepackage{aas_macros_nature}
\usepackage{xfrac}

\usepackage{lineno}

\usepackage{xcolor}
\usepackage[flushleft]{threeparttable}

\definecolor{my_color}{HTML}{3a18b1}

\definecolor{new_color}{HTML}{CF0000}
\definecolor{new_black}{HTML}{000000}

\newcommand{\rsun}{\ensuremath{R_\odot}}
\newcommand{\msun}{\ensuremath{M_\odot}}
\newcommand{\lsun}{\ensuremath{L_\odot}}

\newcommand\arcsec{\mbox{$^{\prime\prime}$}}%

\newcommand{\rearth}{\ensuremath{R_\oplus}}

\newcommand{\teff}{\ensuremath{T_{\rm eff}}}

\newcommand{\vsini}{\ensuremath{V\sin(i_*)}}

\newcommand{\project}[1]{\textsl{#1}}                               
\newcommand{\JWST}{\project{JWST}}

\newcommand{\Gaia}{\project{Gaia}}                            

\newcommand{\Kepler}{\project{Kepler}}  
\newcommand{\kepler}{\project{Kepler}}

\newcommand{\kms}{\ensuremath{\rm km\,s^{-1}}}
\newcommand{\ms}{\ensuremath{\rm m\,s^{-1}}}
\newcommand{\cms}{\ensuremath{\rm cm\,s^{-1}}}

\newcommand{\appropto}{\mathrel{\vcenter{
  \offinterlineskip\halign{\hfil$##$\cr
    \propto\cr\noalign{\kern2pt}\sim\cr\noalign{\kern-2pt}}}}}

\title{A giant planet transiting a 3\,Myr protostar with a misaligned disk}

\begin{document}

\author{Madyson G. Barber$^{1,2}$,
 Andrew W. Mann$^{1}$, 
 Andrew Vanderburg$^{3}$, 
 Daniel Krolikowski$^{4}$, 
 Adam Kraus$^{5}$, 
 Megan Ansdell$^{6}$, 
 Logan Pearce$^{4}$, 
 Gregory N. Mace$^{5}$, 
 Sean M. Andrews$^{7}$, 
 Andrew W. Boyle$^{1}$, 
 Karen A.\ Collins$^{7}$, 
 Matthew De Furio$^{5}$, 
 Diana Dragomir$^{8}$, 
 Catherine Espaillat$^{9}$, 
 Adina D.\ Feinstein$^{10,11}$, 
 Matthew Fields$^{1}$, 
 Daniel Jaffe$^{5}$,  
 Ana Isabel Lopez Murillo$^{1}$, 
 Felipe Murgas$^{12,13}$, 
 Elisabeth R. Newton$^{14}$, 
 Enric Palle$^{12,13}$, 
 Erica Sawczynec$^{5}$,  
 Richard P. Schwarz$^{7}$, 
 Pa Chia Thao$^{1,2}$, 
 Benjamin M. Tofflemire$^{5,15}$, 
 Cristilyn N.\ Watkins$^{7}$, 
 Jon M. Jenkins$^{16}$, 
 David~W.~Latham$^{7}$, 
 George Ricker$^{3}$, 
 Sara Seager$^{3,18,19}$, 
 Roland Vanderspek$^{3}$, 
 Joshua~N.~Winn$^{20}$, 
 David Charbonneau$^{7}$, 
 Zahra~Essack$^{8}$, 
 David~R.~Rodriguez$^{17}$, 
 Avi Shporer$^{3}$, 
 Joseph D. Twicken$^{16,21}$, 
 Jesus Noel Villase{\~ n}or$^{3}$ 
}

\maketitle

\scriptsize
\begin{affiliations}
\item Department of Physics and Astronomy, University of North Carolina at Chapel Hill, Chapel Hill, NC 27599, USA
\item NSF Fellow
\item Department of Physics and Kavli Institute for Astrophysics and Space Research, Massachusetts Institute of Technology, Cambridge, MA 02139, USA
\item Steward Observatory, The University of Arizona, 933 N. Cherry Ave, Tucson, AZ 85721, USA
\item Department of Astronomy, The University of Texas at Austin, Austin, TX 78712, USA
\item NASA Headquarters, 300 E Street SW, Washington, DC 20546, USA
\item Center for Astrophysics \textbar \ Harvard \& Smithsonian, 60 Garden Street, Cambridge, MA 02138, USA
\item Department of Physics and Astronomy, The University of New Mexico, 210 Yale Blvd NE, Albuquerque, NM 87106, USA
\item Institute for Astrophysical Research, Department of Astronomy, Boston University, 725 Commonwealth Avenue, Boston, MA 02215, USA
\item Laboratory for Atmospheric and Space Physics, University of Colorado Boulder, UCB 600, Boulder, CO 80309
\item NHFP Sagan Fellow
\item Instituto de Astrof\'isica de Canarias (IAC), E-38205 La Laguna, Tenerife, Spain
\item Departamento de Astrof\'isica, Universidad de La Laguna (ULL), E-38206 La Laguna, Tenerife, Spain
\item Department of Physics and Astronomy, Dartmouth College, Hanover, NH 03755, USA
\item 51 Pegasi b Fellow
\item NASA Ames Research Center, Moffett Field, CA, 94035, USA
\item Space Telescope Science Institute, 3700 San Martin Drive, Baltimore, MD, 21218, USA
\item Department of Earth, Atmospheric and Planetary Sciences, Massachusetts Institute of Technology, Cambridge, MA 02139, USA
\item Department of Aeronautics and Astronautics, Massachusetts Institute of Technology, Cambridge, MA 02139, USA
\item Department of Astrophysical Sciences, Princeton University, 4 Ivy Lane, Princeton, NJ 08544, USA
\item SETI Institute, Mountain View, CA 94043 USA/NASA Ames Research Center, Moffett Field, CA 94035 USA

\end{affiliations}
\normalsize


\begin{abstract}
\normalfont 
Astronomers have found more than a dozen planets transiting 10-40\, million year old stars \cite{2020Natur.582..497P}, but even younger transiting planets have remained elusive. A possible reason for the lack of such discoveries is that newly formed planets are not yet in a configuration that would be recognized as a transiting planet\cite{Mamajek2012} or cannot exhibit transits because our view is blocked by a protoplanetary disk. However, we now know that many outer disks are warped \cite{Bohn2022}; provided the inner disk is depleted, transiting planets may thus be visible. Here we report the observations of the transiting planet \planetname\ orbiting a 3 Myr, 0.7 $M_\odot$, pre-main sequence star in the Taurus Molecular Cloud. \starname\ hosts a nearly face-on ($i\simeq30^{\circ}$) transitional disk \cite{Espaillat2015} and a wide binary companion. The planet has a period of $8.83$ days, a radius of $10.9$ \rearth\ ($0.97R_J$), and a 95\%-confidence upper limit on its mass of $90M_\oplus$ ($0.3M_J$) from radial velocity measurements, making it a possible precursor of the super-Earths and sub-Neptunes that are commonly found around main-sequence stars. The rotational broadening of the star and the orbit of the wide (4\arcsec, 635 AU) companion are both consistent with edge-on orientations. Thus, all components of the system appear to be aligned except the outer disk; the origin of this misalignment is unclear. Given the rare set of circumstances required to detect a transiting planet at ages when the disk is still present, \planetname\ likely provides a unique window into sub-Neptunes immediately following formation. 


\end{abstract}

\maketitle

\section{Main}\label{sec:main}

\begin{figure}[h]
    \centering
    \includegraphics[width=0.38\textwidth]{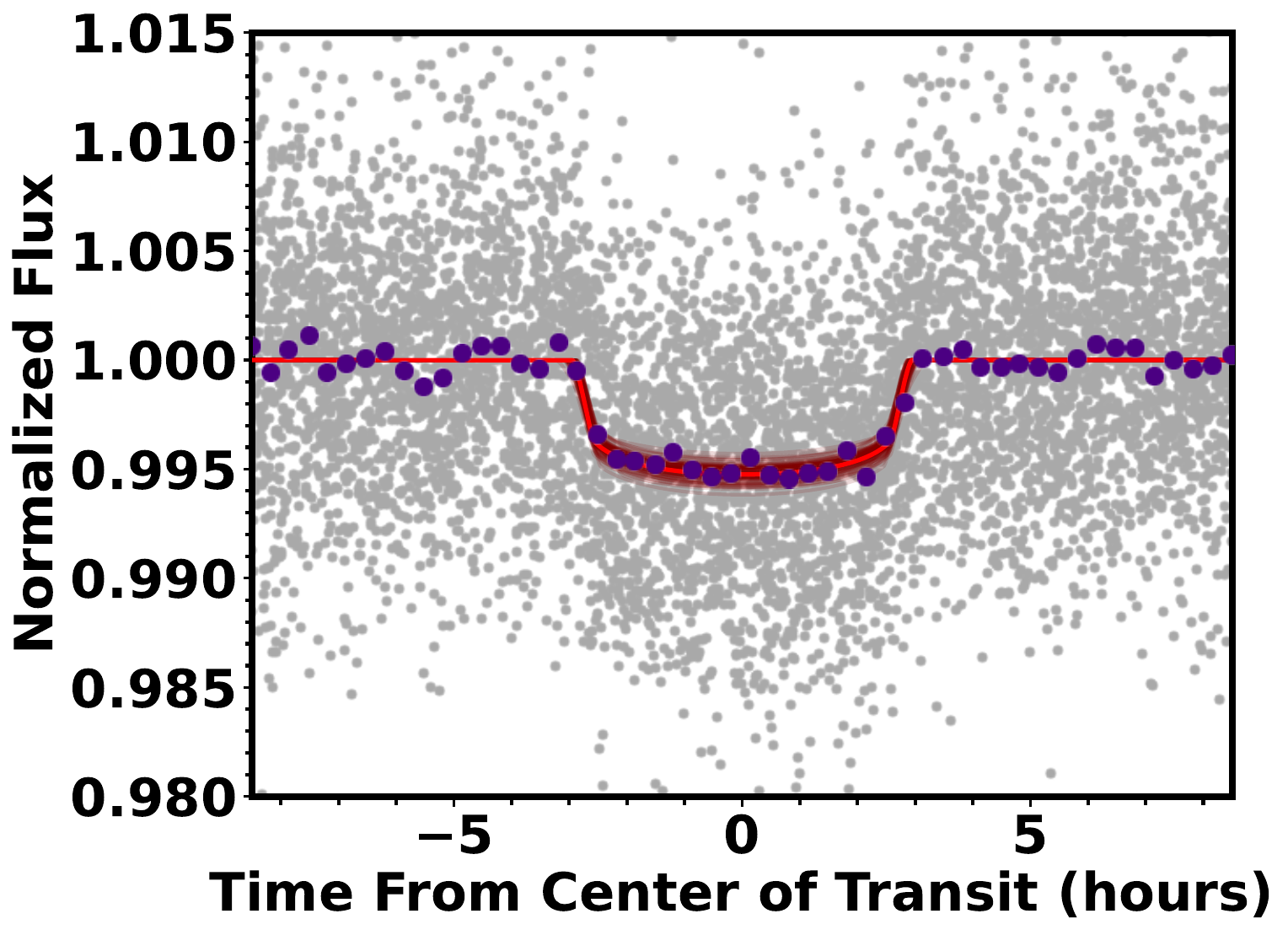}
    \includegraphics[width=0.61\textwidth]{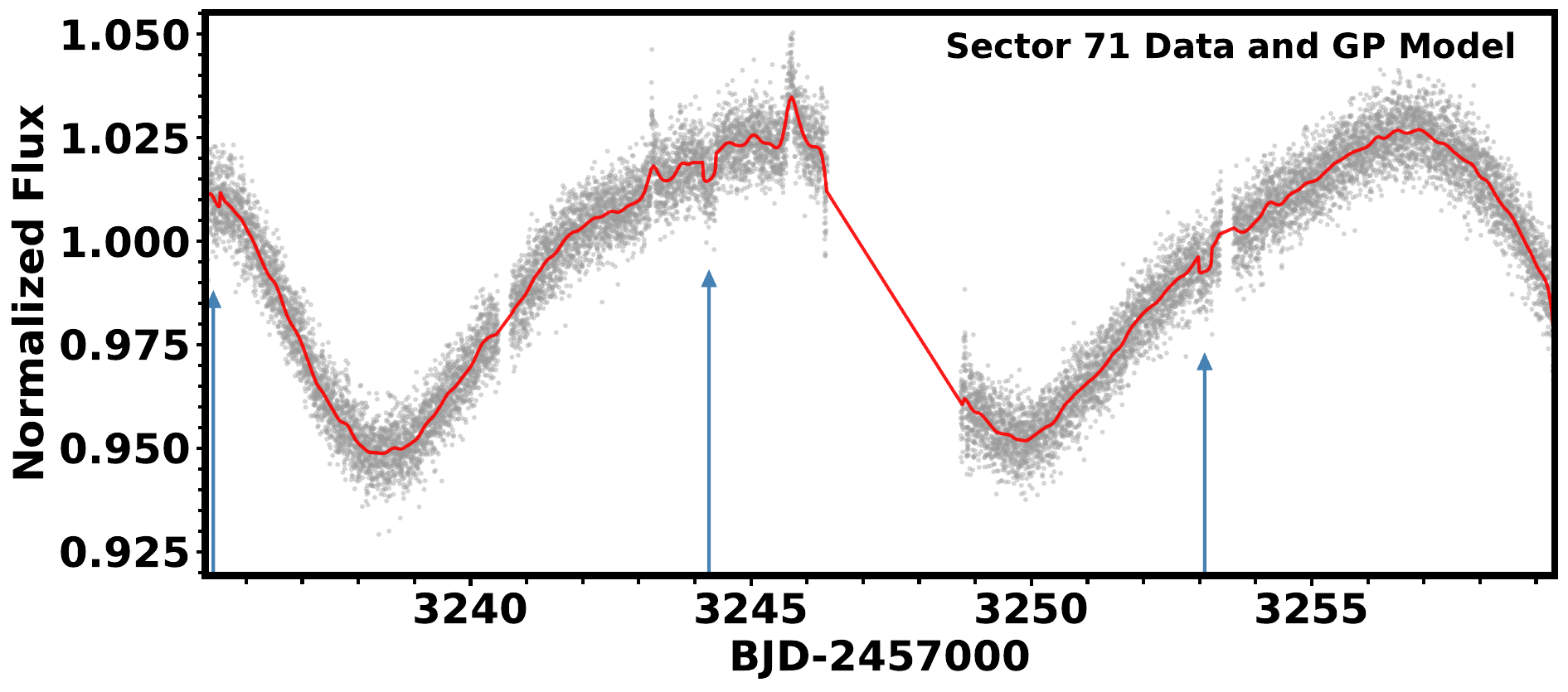}
    \caption{Left: Phase-folded light curve from \tess\ (gray points) binned to 20-minute intervals (purple points). The best-fit transit model is shown as the bright, opaque red line with 100 model fits pulled from the posterior shown as the dark, translucent red lines. The best-fit GP model of the stellar variability was removed from the data and model. Right: Representative sector of the \tess\ light curve (gray points) with the best-fit GP model (red line). Blue arrows indicate transits, which are visible by eye in most cases. In total, across the six sectors, \tess\ observed 18 transits.}
    \label{fig:GPmodel}
\end{figure}

\begin{figure}
    \centering
    \includegraphics[width=0.45\textwidth]{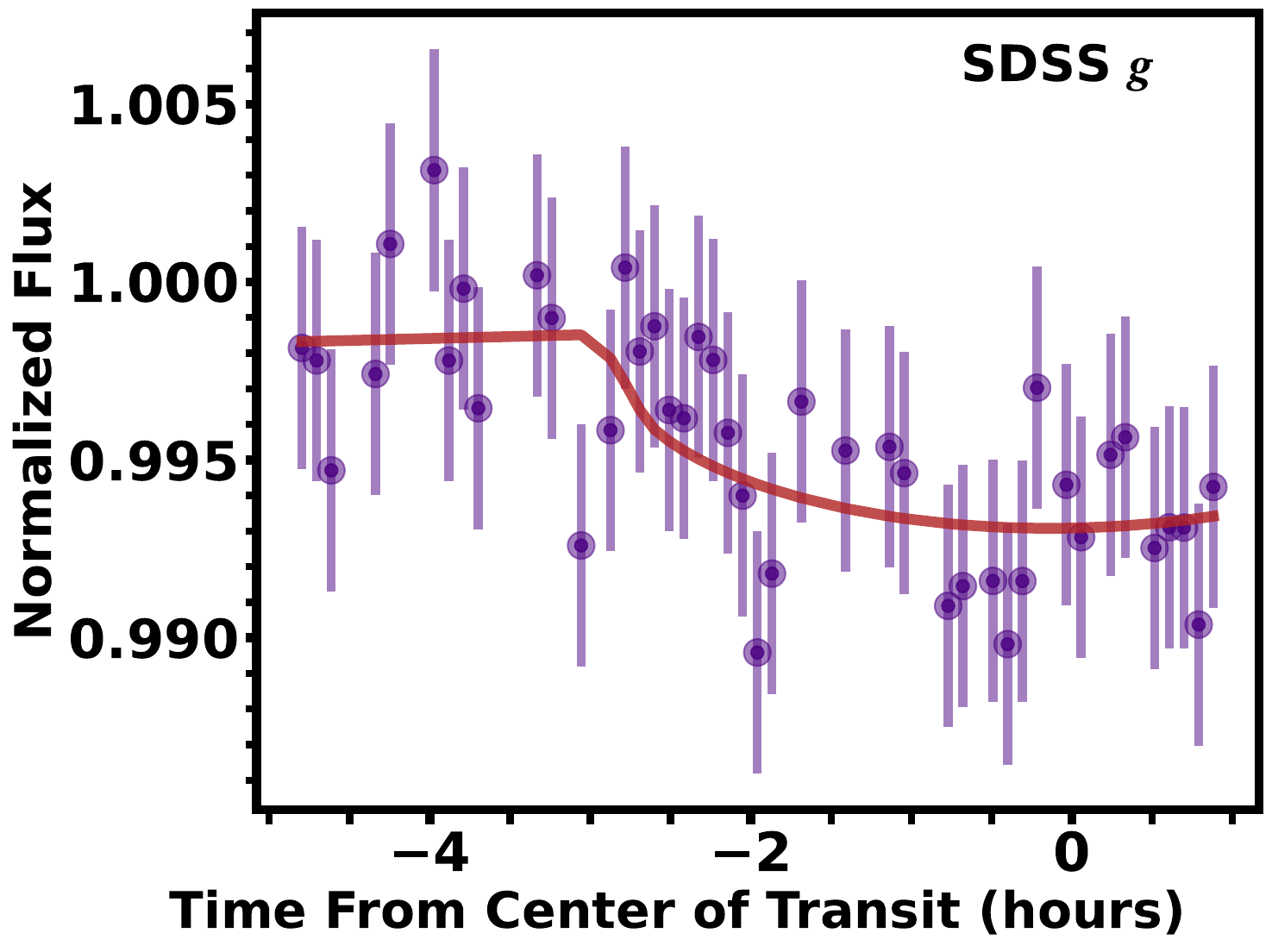}
    \includegraphics[width=0.45\textwidth]{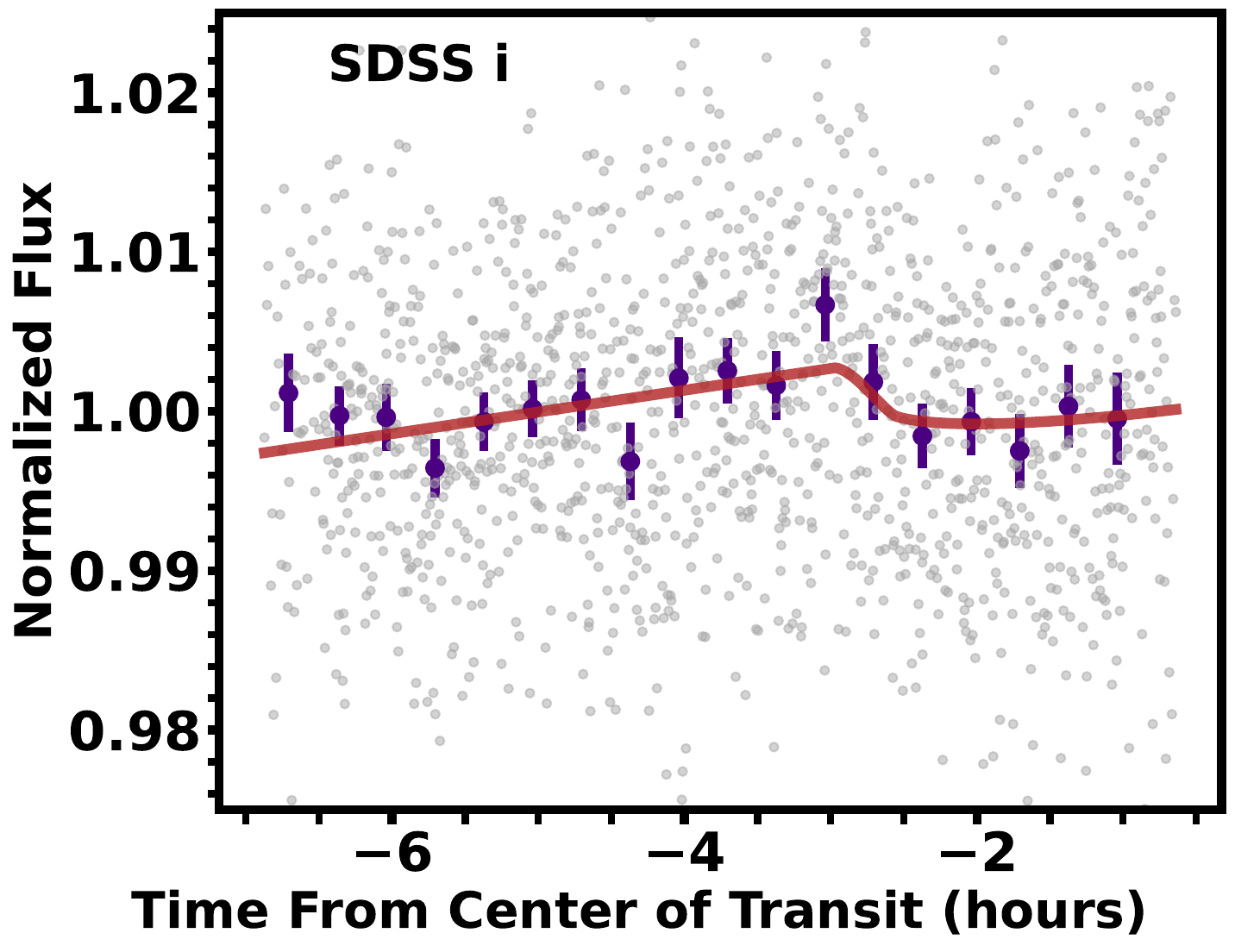}
    \includegraphics[width=0.45\textwidth]{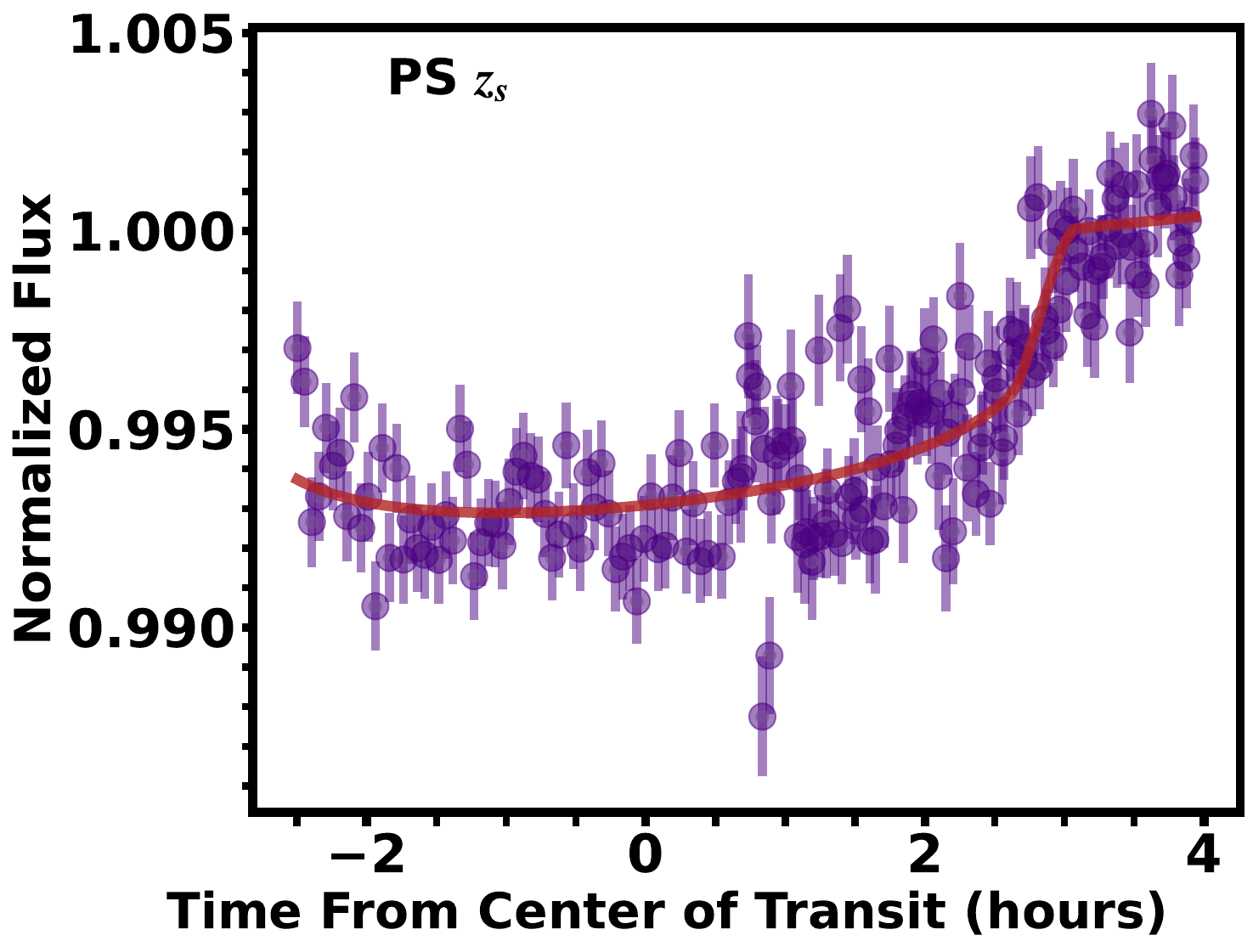}
    \includegraphics[width=0.45\textwidth]{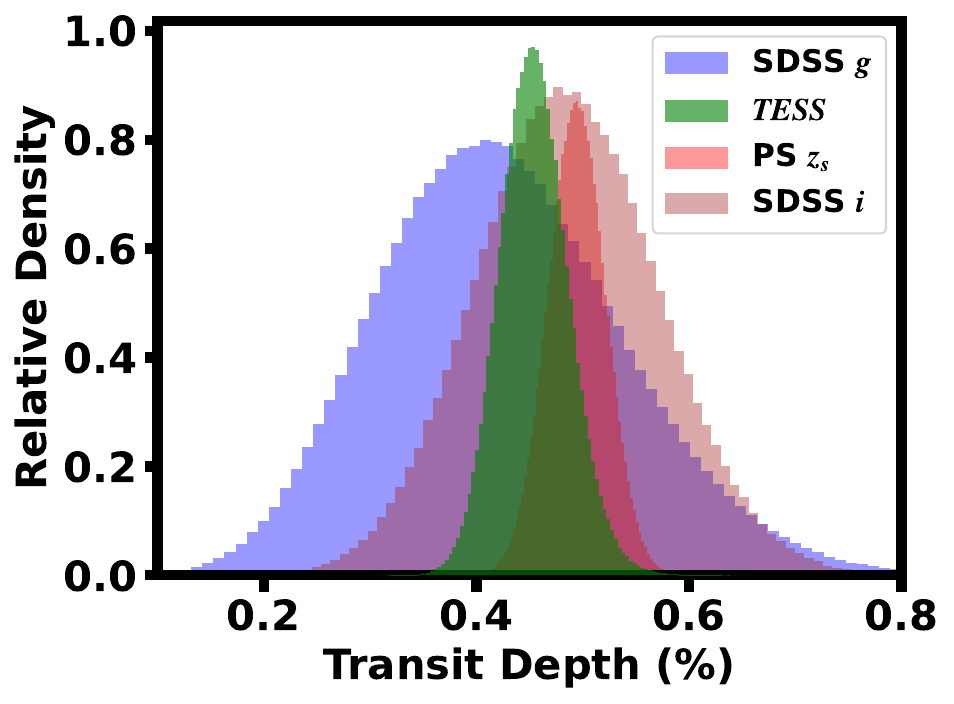}
    \caption{Ground-based follow-up transits observed by LCO. Transits of the same filter have been combined. Binned points are shown for SDSS $i$ for clarity. The models (red lines) are the best-fit transit model with a linear trend. Bottom-right: transit depth posterior from each of the four wavelengths. All calculated transit depths shown in Extended Table \ref{tab:transitDepths}. }
    \label{fig:LCOtransits}
\end{figure}

\starname\ was first observed by NASA's Transiting Exoplanet Survey Satellite (\tess) in November 2019, and since then, \tess\ has detected 18 transits of \planetname\ across six sectors (Figure~\ref{fig:GPmodel}). Follow-up observations using the Las Cumbres Observatory (LCO) 1-meter telescopes confirmed with higher angular resolution that \starname\ is the source of the photometric variations seen by \tess\ and has a consistent depth at wavelengths $0.5-0.9$\um\ (Figure~\ref{fig:LCOtransits}). Our analysis of the light curve indicates that the planet orbits its star every 8.83 days and has a radius of $0.97\pm0.06$~$R_J$. The multi-color observations of transits, radial velocities (RVs), and multi-epoch high-resolution imaging combine to rule out all false positive scenarios (see Methods).

\starname\ is an unambiguous member of the Taurus-Auriga star-forming region. Taurus-Auriga itself is made up of at least 17 distinct subpopulations ranging in age from 1.3 to 15 Myr. \starname\ is a member of the D4-North subpopulation, with an estimated age of $2.49^{+0.35}_{-0.34}$ Myr \cite{Krolikowski2021}. By comparing the spectra, photometry, and \gaia\ parallax to a grid of evolutionary models \cite{PARSEC}, we determined an age of $3.3^{+0.6}_{-0.5}$ Myr for the host star, consistent with the group age.

This makes \planetname\ the youngest transiting planet discovered to date by a factor of $\simeq3$. There are similar-age directly imaged systems \cite{kraus2014a, Fontanive2020}, but those reside near the boundary between brown dwarfs and planets. Radial velocity surveys have also uncovered similar-aged non-transiting planet candidates \cite{2016ApJ...826..206J, Donati2016}, but those remain controversial \cite{Donati2020, 2020A&A...642A.133D}. Even if confirmed, those planets are $>1M_J$, making them progenitors of brown dwarfs or hot Jupiters.
 
The analysis of archival and new RVs sets an upper limit of $<0.3\,M_J$ on the planet's mass. Planets younger than $\sim$100\,Myr are expected to have inflated radii and will shrink over time\cite{Fortney2007}. Gas-giant planet progenitors ($\gtrsim1M_J$) are expected to be 1.4-2$R_J$ for their first 20-30\,Myr\cite{Spiegel2012}. Observational data shows 10-700\,Myr planets are larger and have a lower density than their older counterparts \cite{Fernandes2023, Vach2024}. Lastly, small planets are intrinsically more common, particularly around lower-mass stars, so the likelihood of finding a hot Jupiter orbiting a 0.7\msun\ star is small \cite{Fressin2013}. Thus observational and statistical data indicate \planetname\ is not a true giant, but instead a progenitor of a sub-Saturn ($4-7R_\oplus$) or even a sub-Neptune ($\simeq1.5-4R_\oplus$).

\begin{figure}
\begin{centering}
\includegraphics[width=8.5cm]{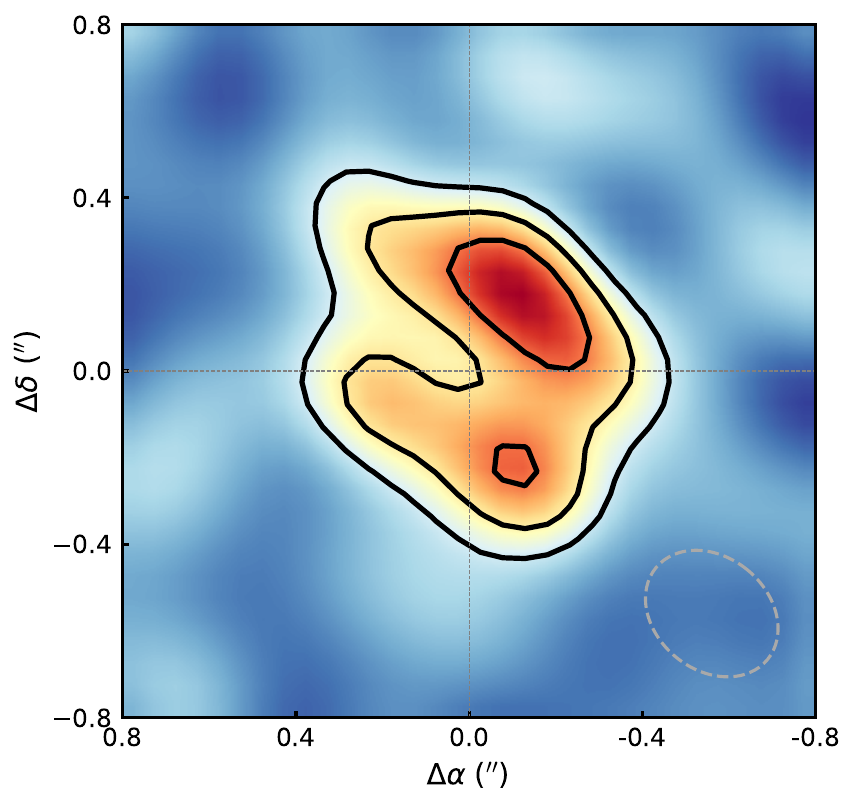}
\caption{Archival \small SMA 880~$\mu$m continuum emission of IRAS 04125+2902\cite{Espaillat2015}. The image is centered on the estimated disk center and is scaled so that reddest color is equal to the peak flux density.
Contours at $2\sigma$, $4\sigma$, and $6\sigma$ are shown by the black solid lines. The synthesized beam is illustrated by the dashed gray ellipse drawn in the lower right corner. The irregular shape is likely due to the resolution of the data rather than asymmetries of the disk.}
\label{fig:disk}
\end{centering}
\end{figure}

Pebble accretion models can form and migrate a solid core of $\sim$tens of Earth masses to $<0.2$AU within 2-3\,Myr\cite{Johansen2017}. The existence of \planetname\ is not surprising in light of these models and provides observational evidence that planets can form and migrate on such a short timescale. It also demonstrates that a planet at this age can be a spherical object with a sharply defined photosphere (forming a trapezoid-like transit) rather than a cloud of dust and gas or a still-forming ring-like system \cite{Mamajek2012}.

Protoplanetary disks dissipate from the inside out, creating large inner dust cavities, often tens of AU in diameter. Disks with such cavities are known as transition disks \cite{Espaillat2014}. Stars with transition disks have long been targeted by direct imaging planet searches, both because the cavity may have been cleared by a giant planet\cite{Zhu2011} and because the region is optically thin enough to detect such a planet\cite{Haffert2019}. \starname\ shows clear evidence for a transition disk (Figure \ref{fig:disk}), based on its spectral energy distribution and on submillimeter observations of the outer disk \cite{Espaillat2015}, but no directly-imaged planets inside the gap have been found to date \cite{Ruiz-Rodriguez2016}.

It was previously thought that the detection of a transiting planet in a disk-bearing system was improbable. A transiting planet must have a high orbital inclination (edge-on) and is assumed to have formed within the (sky-projected) plane of the disk, but an optically thick edge-on disk would also block observations of the host star at wavelengths of typical transit observations. However, the outer disk surrounding \starname\ is close to face-on when observed from Earth \cite{Espaillat2015} and has a depleted inner disk out to $\simeq$20\,AU.

\begin{figure}
    \centering
    \includegraphics[width=0.8\textwidth]{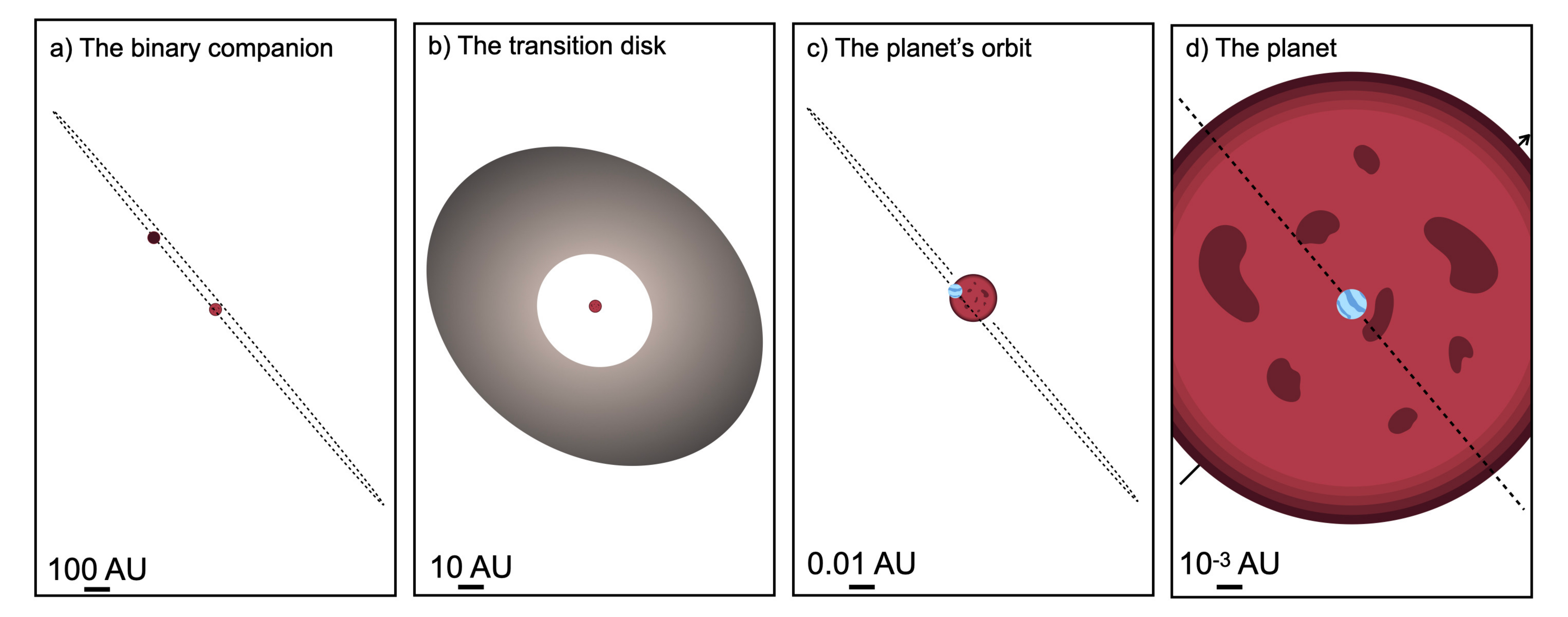}
    \caption{Schematic of the system showing the stellar orbit (a), the disk (b), the planet orbit (c), and the planet (d). The scale is correct within each panel, and the inclination angle is preserved between panels. Note that the ascending node ($\Omega$) is not known for the star or planet, but are drawn to match the binary. }
    \label{fig:sketch}
\end{figure}

The stellar rotational axis is consistent with that of the planet's orbital axis, but nearly perpendicular to the disk orientation (Figure \ref{fig:sketch}, see Methods). The star has a 4\arcsec\ binary companion, which could explain the disk misalignment. However, the orbit of the binary is also edge-on (see Methods), as has been seen in some other planetary systems with binary companions \cite{THYMEI}. The lowest energy configuration is such that the system components should be aligned, suggesting the companion-disk interactions should preferentially push the disk {\it into} alignment with the binary's orbit.

The origin of the misaligned outer disk remains unclear. One possibility is that the disk components were initially aligned with each other, and that the planet became misaligned during migration. Current models for forming close-in sub-Neptunes tend to include migration of pebbles and/or planetesimals through the protoplanetary disk \cite{Johansen2017}, which produce planets aligned with their parent disk. Another massive object in the system is likely required to move a planet into an orbit inclined from the disk that formed it, though there is no evidence of such an object in the system.  

The alternative is that the outer disk became warped through a process distinct from interactions with the binary companion. Simulations of star-forming regions suggest outer disk warps could be common due to interactions between the early disk and the surrounding environment\cite{Bate2018}, such as infalling material from the surrounding cloud \cite{Kuffmeier2021}. Observational studies have also found increasing evidence of disk warps \cite{Casassus2018, Kraus2020disk, Hurt2023}, although others have found general alignment \cite{Davies2019}. However, the frequency of outer disk warps at such young ages is not known. Thus, whether \starname\ is a special case or one of a larger population of systems with broken/warped disks \cite{Benisty2018} is still unclear.

Given its close proximity to Earth (160 pc) and rare configuration, the \starname\ system is a powerful environment for understanding early formation and migration. Rossiter–McLaughlin measurements and future ALMA data can be used to further constrain the system alignment and disk inclination. Due to the large radius and relatively low mass, we can conclude that \planetname\ likely has a large extended atmosphere, ideal for follow-up with \JWST. \planetname\ motivates further efforts to search the youngest stellar clusters for more of these infant planets.

\newpage
\begin{methods}

\renewcommand{\figurename}{Extended Data Figure}
\renewcommand{\tablename}{Extended Data Table}
\setcounter{table}{0}  
\setcounter{figure}{0}

\subsection{\tess\ observation and identification of the planet:}

\starname\ (TIC 56658270) was first observed by the Transiting Exoplanet Survey Satellite (\tess) in Sector 19, from 2019 Nov 28 to Dec 23. The target was re-observed by \tess\ in Sector 43 (2021 Sep 16 - Oct 11), Sector 44 (2021 Oct 12 - Nov 5), Sector 59 (2022 Nov 26 - Dec 23), Sector 70 (2023 Sep 20 - Oct 16) and Sector 71 (2023 Oct 17 - Nov 11). For Sectors 43, 44, 59, 70, and 71, the target was pre-selected for 2-minute cadence light curves produced by the Science Processing Operations Center (SPOC)\cite{Jenkins:2016}. Sector 19 was observed only in the 30-minute cadence Full Frame Images (FFIs).

As part of our survey for planets transiting stars in the youngest stellar associations, we searched the \tess\ light curves of all high-probability members of the Taurus-Auriga association\cite{Krolikowski2021}. We used a custom extraction pipeline starting from \tess\ FFIs and SPOC Simple Aperture Photometry (SAP) \cite{Twicken2010, Morris2020SPOC}. We extracted the light curves and applied our custom systematic corrections following the procedure in reference \cite{Vanderburg2019}. To briefly summarize, we started with 30x30 pixel cutouts of \tess\ FFIs. Using the calibrated pixel files, we tested a range of circular and irregular aperture shapes, adopting the one that minimized the photometric scatter and nearby-star contamination. The resulting light curves were then corrected with a linear model including a regularly-spaced B-Spline, several moments of the distribution of the spacecraft quaternion time series measurements within each exposure, the robust mean of pixels in the background of the FFI cutouts, and seven co-trending basis vectors from the SPOC Presearch Data Conditioning \cite{Stumpe2012, Smith2012, Stumpe2014} single-band and band 3 (fast timescale) flux time series correction with the largest eigenvalues. The SPOC SAP curves were fit with a similar procedure, but with additional basis vectors and a high-pass filter.

We searched for planets using the \texttt{Notch} pipeline\cite{Rizzuto2017}. From the original code, we updated the box-least squares (BLS) search to the one implemented in \texttt{astropy} \cite{AstropyCollaboration2018}, and changed the period grid spacing used. The search recovered a signal at 8.8\,days with a BLS SNR of 42. The resulting phase-folded transit showed a sharp ingress/egress with a flat bottom consistent with a planetary object.

We performed a suite of initial quality checks on the detection, including retrieval of the planet using multiple light curve pipelines \cite{Bouma2019, Hattori2021}, checking depth and duration consistency with varying aperture selection and across all six sectors. Passing these checks triggered our follow-up with a suite of high-resolution imaging, multiwavelength transits, and high-resolution spectroscopy, and checking for corresponding archival data which we detail below. 

The \tess\ Science Office vetting team independently identified the signal and upgraded the system to TOI-6963 on April 3 2024. 

\subsection{High-Resolution Imaging:}

\starname\ has archival high-resolution imaging spanning November 2009 to November 2016. This includes deep $H$-band imaging from the Strategic Exploration of Exoplanets and Disks with Subaru (SEEDS)\cite{Uyama2017}, as well as $L_p$-band imaging\cite{Wallace2020}, $K'$-band imaging, and $K'$-band non-redundant aperture masking (NRM) \cite{Ruiz-Rodriguez2016} with NIRC2 on Keck-II. 

To provide a more recent epoch, we observed \starname\ on 2023 Dec 5 with $K'$-band adaptive optics imaging using NIRC2 on Keck-II. These observations were taken, reduced, and analyzed following prior strategies optimized for identifying stellar companions to \kepler\ planet hosts \cite{Kraus2016a}. 

We summarize the resulting limits from the high-resolution data in Extended Figure~\ref{fig:patientImaging} by band and time.

\subsection{Ground-based Transits:}

We observed five transits of \starname\ with the Las Cumbres Observatory telescope network (LCO)\cite{Brown13}. These were observed with Sinistro cameras, with a pixel scale of 0.389\arcsec\,pixel$^{-1}$. On 2023 December 9 and 18, we observed egress using the Pan-STARRS $z_s$ filter; on 2024 January 14, we observed an egress with the Sloan $g'$ filter; and on 2024 Jan 22, we observed ingress simultaneously with two telescopes both using the SDSS $i'$ filter (we analyzed each separately). The exposure times were 300, 150, and 3\,s for $g'$, $z_s$, and $i'$ observations, respectively. Due to the long transit duration, all were partial transits. 

All images were initially reduced by the standard LCOGT {\tt BANZAI} pipeline\cite{McCully18}. We tested extracting photometry from {\tt BANZAI}, {\tt AstroImageJ}\cite{Collins:2017}, and {\tt Photutils}, varying sets of between 10 and 15 comparison stars, and varying aperture sizes (diameter of 1-6\arcsec). We recovered the transit with consistent depth across all methods.

\subsection{Radial Velocities of \starname:}

We retrieved two sets of spectra of \starname\ taken with the Immersion Grating Infrared Spectrometer (IGRINS)\cite{Park2014} on 2015 Jan 27 and 2017 Dec 02. These observations were taken as part of a survey of young stellar objects in nearby star-forming regions\cite{Lopez-Valdivia2023}. We reduced the IGRINS spectra with the publicly available IGRINS pipeline package \cite{IGRINS_plp} and extracted RVs using the \texttt{IGRINS RV} code \cite{IGRINS_RV}. The RVs were consistent with no variation over a 3-year baseline. 

We drew four additional RV measurements from the seventeenth APOGEE data release\cite{Abdurro2022}. The RVs span 2019 January 20 to 2020 January 6, and are consistent with no variation over this period. 

We obtained six new observations of \starname\ with the Habitable-zone Planet Finder (HPF)\cite{Mahadevan2012,mahadevan2014} on the 10-m Hobby-Eberly Telescope at McDonald Observatory. HPF is a high resolution ($R\sim55000$), fiber-fed\cite{Kanodia2018}, stabilized\cite{stefansson2016} spectrograph covering $8100-12700$~\AA. HPF has a laser frequency comb (LFC) to achieve extremely high quality wavelength calibration, exhibiting $\sim20$~\cms\ instrumental calibration precision\cite{Metcalf2019}. 

Each HPF observation was composed of three sequential 600-s exposures to increase signal and average over high-frequency, oscillation-driven stellar RV variation. To minimize the effect of stellar activity-driven RV noise, we scheduled observations within one stellar rotation (11\,days) while sampling at orbital phases of maximum velocity expected from the planet. We bracketed each observation with LFC exposures to monitor the instrumental velocity drift. 

The HPF data were reduced using the HxRGproc toolbox\cite{Ninan2018, Kaplan2019, Metcalf2019}. We corrected the derived wavelength solution for barycentric motion at the flux-weighted exposure midpoint using \texttt{barycorrpy}\cite{Wright2014}. We measured RVs using broadening functions\cite{Rucinski1992,Rucinski2002} computed using the \texttt{saphires} Python package\cite{Tofflemire2019,saphires}, which compares each individual spectrum to a template. We used an empirical template of \starname\ generated by stacking all HPF observations, and measured the RV by fitting a Gaussian to the combined broadening function for each HPF exposure and binned the three RVs comprising each observation. We repeat this procedure using a Phoenix model spectrum\cite{2013A&A...553A...6H} to provide RVs on the same scale as the IGRINS and APOGEE RVs, because comparison to an empirical template offers precise but relative RVs.

All RVs are summarized in Extended Table~\ref{tab:rvs} and Extended Figure~\ref{fig:rvs}

\begin{figure}
    \centering
    \includegraphics[width=0.8\textwidth]{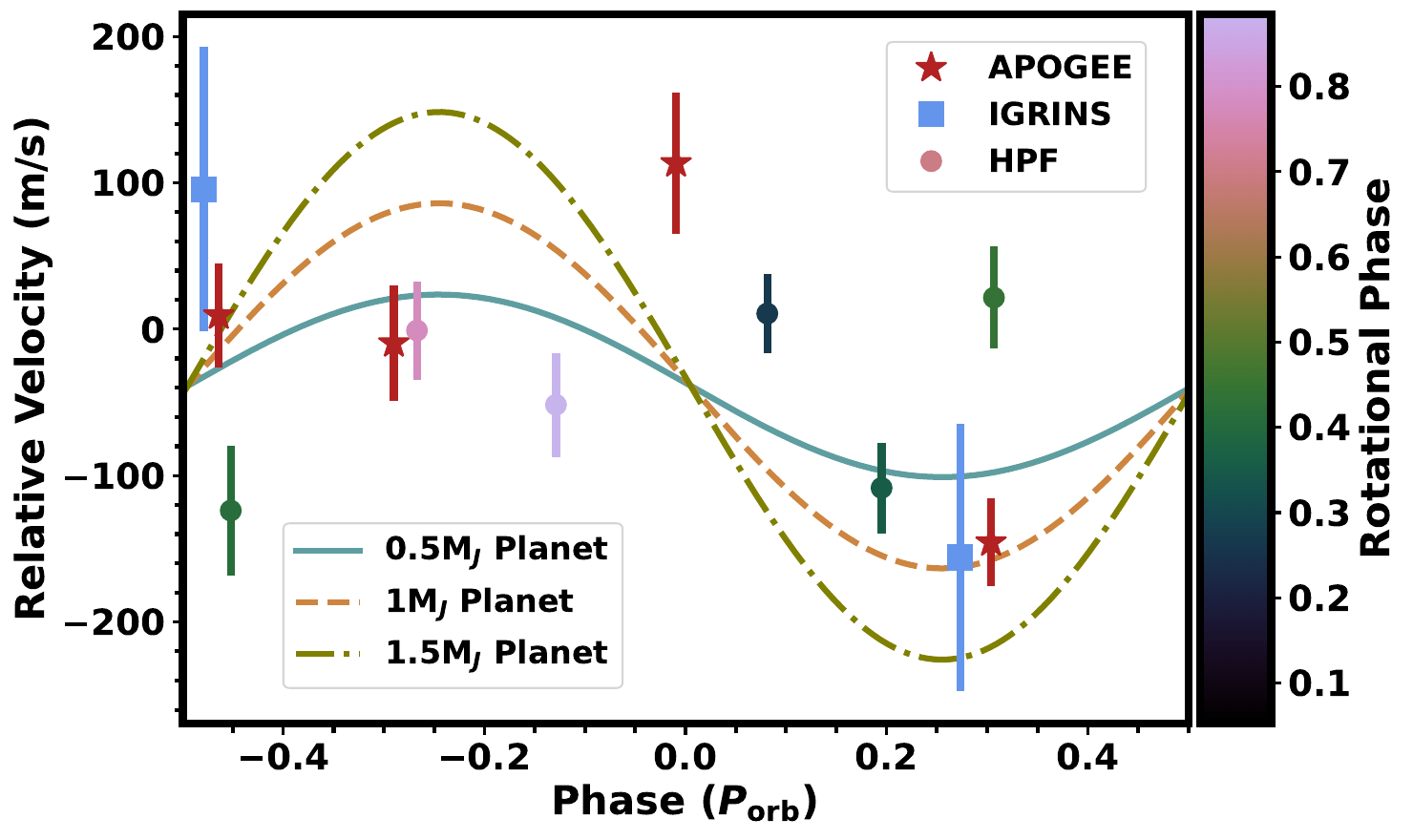}
    \caption{Radial velocities of \starname\ phased to the period of the planet. RVs from APOGEE are shown as red stars, those from IGRINS are shown as blue squares, and those from HPF are shown as circles colored by the stellar rotation phase (with an arbitrary zero point). We show the expected signals from a 0.5, 1, and 1.5 $M_J$ companion. The RVs rule out any companion of stellar mass, and the best-fit of the signal from the HPF measurements puts an upper limit of the planet mass at $<0.3M_J$ at 95\%. }
    \label{fig:rvs}
\end{figure}

\begin{table}
    \centering
    \begin{tabular}{cccccccccc}
    \hline
    \hline
    Epoch & Absolute RV & $\sigma_{Abs.\,RV} $ & Relative RV & $\sigma_{Rel.\,RV}$ & Source \\
    (BJD) & (\kms) & (\kms) & (\ms) & (\ms) &  \\
    \hline
    2457050.603 & 17.174 & 0.097 & 95.7 & 97.3 & IGRINS \\
    2458090.937 & 16.923 & 0.091 & -155.9 & 91.4 & IGRINS \\
    2458499.666 & 17.088 & 0.036 & 9.5 & 35.6 & APOGEE \\
    2458503.684 & 17.192 & 0.048 & 113.5 & 48.3 & APOGEE \\
    2458744.995 & 16.933 & 0.030 & -145.5 & 30.2 & APOGEE \\
    2458854.603 & 17.069 & 0.040 & -9.5 & 39.5 & APOGEE \\
    2460306.817 & 16.91 & 0.10 & 10.7 & 27.0  & HPF \\
    2460307.821 & 16.87 & 0.11 & -108.4 & 31.0 & HPF \\
    2460308.808 & 17.00 & 0.12 & 21.6 & 34.9 & HPF \\
    2460312.574 & 16.93 & 0.11 & -0.9 & 33.6 & HPF \\
    2460313.795 & 16.98 & 0.13 & -51.6 & 35.5 & HPF \\
    2460319.773 & 16.80 & 0.16 & -123.9 & 44.3  & HPF \\
    \hline
    \end{tabular}
    \caption{Radial velocities of \starname. For IGRINS and APOGEE, the relative RVs are normalized to each other. For HPF, the relative RVs are the reduced values, as described in text.}
    \label{tab:rvs}
\end{table}

\subsection{Host star properties:} 

We determined $R_*$, $L_*$, and \teff\ by simultaneously fitting the available photometry and the archival optical spectrum \cite{Herczeg2014} as done for prior work on young planet-hosts\cite{Mann2016b}. We excluded data past 8\um, which includes emission from the disk. There was no evidence of significant contribution of the disk blueward of this in the SED fit. After accounting for stellar variability, the best-fit template reproduced the spectrum extremely well ($\chi^2_\nu=1.1$, Extended Figure \ref{fig:sed}).

We estimated the bolometric flux (\fbol) from the integrated spectrum, the luminosity from \fbol\ and the \gaia\ parallax, and $R_*$ from the Stefan-Boltzmann relation. The final fit yielded \teff=$3912\pm72$\,K, $L_*=0.466\pm0.041R_\odot$, $A_V=2.71\pm0.20$, and $R_*=1.48\pm0.07R_\odot$. These values are consistent to those independently measured in reference \cite{Espaillat2015}, and $A_V$ is consistent with findings in references \cite{Furlan2011, Andrews:2013yq, Espaillat2015}.

\begin{figure}
    \centering
    \includegraphics[width=0.51\textwidth]{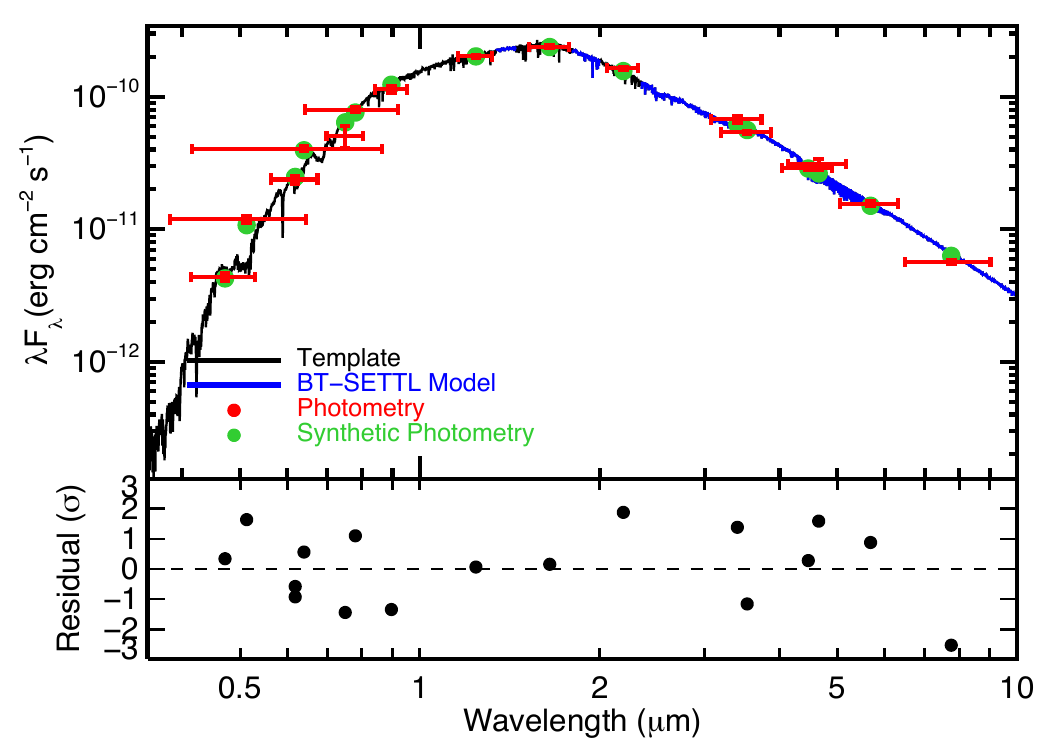}
    \includegraphics[width=0.41\textwidth]{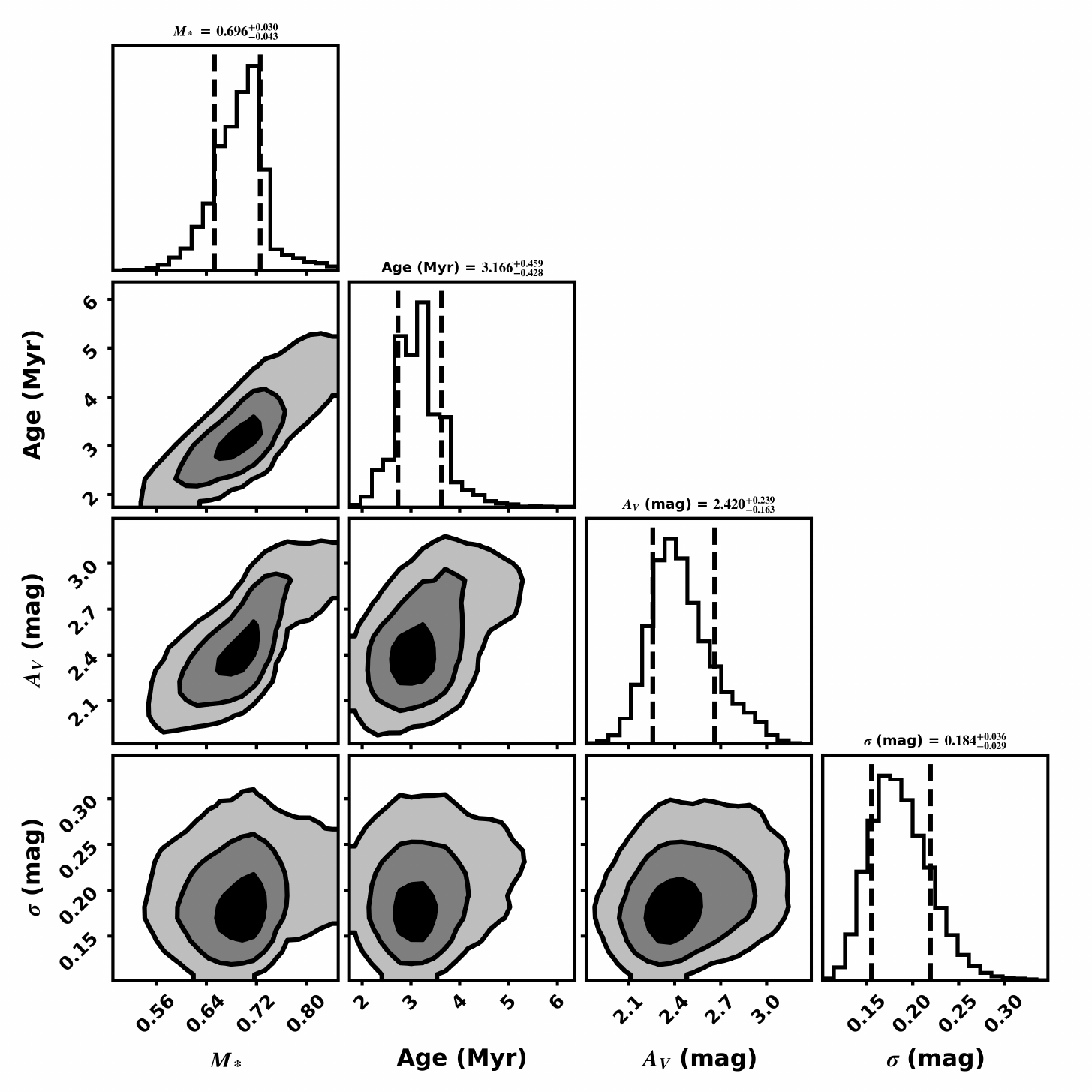}
    \caption{Left: example fit to the SED, showing a template spectrum (black), BT-SETTl models (blue), observed photometry (red) and the synthetic photometry (green). The vertical errors are the photometric uncertainties, while the horizonal error bars indicate the width of the filter. The bottom panel shows the residuals in standard deviations. Right: Corner plot of the PARSECv1.2S model grid MCMC fit. }
    \label{fig:sed}
\end{figure}

To estimate $M_*$ and age, we compared photometry and distance to the PARSECv1.2S model grid \cite{PARSEC} within an MCMC framework\cite{Foreman-Mackey2013}. As with the SED fit, we excluded photometry past 8\um. The free parameters were the stellar mass ($M_*$), age, extinction ($A_V$), and underestimated uncertainty ($f$). We placed a Gaussian prior on $A_V$ based on a fit of the optical spectrum with a stellar template, but other parameters evolved under uniform priors. This yielded a mass of $0.70\pm0.04M_\odot$ and an age of $3.3^{+0.6}_{-0.5}$\,Myr (Extended Figure \ref{fig:sed}). This age is older than but consistent with the parent Taurus-Auriga sub-group \cite{Krolikowski2021}. 

We tried fitting the primary and 4\arcsec\ companion simultaneously, assuming the same age. Because the companion has less and lower-quality photometry, the fit was almost unchanged ($3.6\pm0.6$\,Myr) from the primary-only fit. However, this provided an mass estimate $M = 0.17\pm0.04 M_\odot$ for the companion we use in our binary orbit fit. We summarize the host and companion stellar properties in Extended Table \ref{tab:stellarParameters}.

\subsection{Star-planet alignment:}

We estimated the inclination of the host star ($i_*$) using the combination of the stellar rotation period (\prot), rotational broadening (\vsini), and the stellar radius ($R_*$).

We measured \prot\ by running a Generalized Lomb Scargle (GLS) periodogram\cite{wbhatti_astrobase} on the \tess\ light curves. We used the simple-aperture photometry (SAP) fluxes, which we median-normalized to unity within a given sector, then stitched all sectors together. For the \prot\ uncertainty, we used the scatter in \prot\ determinations between sectors and the stitched curve. We find a final rotation period \prot\ $= 11.31 \pm 0.06$ days.

To estimate rotational broadening (\vsini) we cross-correlated the IGRINS spectra with artificially broadened IGRINS spectra of slowly-rotating stars with similar \teff\ to \starname. The adopted \vsini\ was the broadening that gave the maximum cross-correlation\cite{Kesseli2018}. To estimate uncertainties, we repeated this process over multiple orders and with multiple templates, yielding a \vsini\ of $7.1\pm0.5$\kms, consistent with earlier determinations \cite{Lopez-Valdivia2023}.

Using the $R_*$ and \prot\ values above gave an equatorial velocity of 6.6$\pm$0.3\kms, which we combined with our \vsini\ estimate to set limits on the the stellar inclination ($i_*$)\cite{Masuda2020}. This yielded an lower limit on $i_*$ of 78$^\circ$ at 1$\sigma$ and 68$^\circ$ at 2$\sigma$. The resulting inclination is consistent with that of the transiting planet within 1$\sigma$. Despite the inclinations agreeing, the ascending nodes ($\Omega$, the sky-projected angles) are unconstrained, and therefore the host and planet are not necessarily truly aligned.

\subsection{Star-binary alignment:}

\gaia\ detected a star $\simeq 4\arcsec$ away (\companionname) with matching parallax and proper motions. This corresponds to a sky-projected separation of 635~AU. We recovered this source in the three epochs from NIRC2/Keck. In each epoch, we measured the relative astrometry by fitting the PSF with an empirical template built from other images taken in the same night \cite{Kraus2016a}. We applied a $-0.062\deg $ offset to the NIRC2 position angles to put them on the same scale as the \gaia\ astrometry.

A wide binary companion can explain the presence of a warped disk depending on the binary orbit \cite{Lubow2000}. To investigate this, we fit the orbit of \companionname\ relative to \starname\ using \texttt{lofti} \cite{Pearce2020}. \texttt{lofti} uses the orbit fitting rejection sampling algorithm of \texttt{OFTI} \cite{Blunt2017} with Gaia proper motions and our Keck astrometry as observational constraints. 

Extended Figure \ref{fig:companionInclination} shows posterior distributions for the semi-major axis, period, and inclination resulting from our binary orbit fit. Due to the astrometric points covering only a small fraction of the orbit, the eccentricity is unconstrained \cite{FerrerChavez2021BiasesOrbitFitting}. Both the \gaia\ and Keck astrometry are consistent with purely radial motion; hence, the inclination ($i_c$) is constrained to 94.5$^{+10.9}_{-4.7}$ degrees. This is consistent with the edge-on orbit of the star-planet system (within 1$\sigma$).

\subsection{Disk-star alignment and disk properties:}

\starname\ is known to harbor a transitional disk, as evident from the SED peak at $\simeq$100\um\ and a central hole seen in submm/mm imaging \cite{Espaillat2015}. 

To confirm a previous estimate of disk inclination ($i_d$) \cite{Espaillat2015}, we reanalyzed the Submillimeter Array (SMA) data. We fit a simple elliptical Gaussian to the visibility data with the {\it uvmodelfit} task in CASA (version 6.5.6.22). Using CASA's {\it importuvfits} task, we converted the SMA observations in UVFITS format to a CASA visibility data set. The {\it uvmodelfit} elliptical Gaussian model has six free parameters: integrated flux density ($F_{\rm cont}$), FWHM along the major axis ($a$), aspect ratio of the axes ($r$), position angle (PA), right ascension offset from the phase center ($\Delta\alpha$), and declination offset from the phase center ($\Delta\delta$). We scaled the uncertainties on the fitted parameters by the square root of the reduced $\chi^{2}$ value of the fit. We found $r=0.856\pm0.077$, which implies an outer disk inclination $31\pm12^{\circ}$, consistent with the prior determination of $\lesssim30^{\circ}$\cite{Espaillat2015}. We show the image extracted from the visibilities in Figure~\ref{fig:disk}. The disk inclination disagrees with the star-planet-binary system by $\sim4\sigma$.

{\bf Planet Properties}

\subsection{Transit analysis:}

For the transit, we use our FFI-extracted light curve for Sector 19 and the 2-minute systematic-corrected SPOC PDCSAP light curves for Sectors 43, 44, 59, 70, and 71. We remove one transit in Sector 43 (expected $T_0$ $\simeq 2484.5$ BTJD) as the transit occurs close to the \tess\ data edge. We remove a flare at $\sim 2519.95$ BTJD, which marginally interrupts the egress of a transit in Sector 44 (expected $T_0$ $\simeq 2519.8$ BTJD).

We fit the systematics-corrected \tess\ photometry using \texttt{MISTTBORN} (MCMC Interface for Synthesis of Transits, Tomography, Binaries, and Others of a Relevant Nature) code\cite{Mann2016a,MISTTBORN}. \texttt{MISTTBORN} utilizes \texttt{BATMAN} \cite{Kreidberg2015} to generate the light curve models, \texttt{emcee} \cite{Foreman-Mackey2013} to explore the parameter space, and \texttt{celerite} \cite{celerite} to model the stellar variability with a Gaussian Process (GP).

We fit 12 parameters in total. Five are the common transit parameters: time of periastron ($T_0$), planet orbital period ($P$), planet-to-star radius ratio ($R_p/R_*$), impact parameter ($b$), and stellar density ($\rho_*$). We also fit two quadratic limb darkening coefficients ($q_1$ and $q_2$) following the triangular sampling prescription\cite{Kipping2013}.

The last five parameters were used to model the GP in the \texttt{celerite} code. We used two simple harmonic oscillators (SHOMixture) to model the stellar variability. The GP consisted of five parameters: the natural logarithm of the GP period ($\ln{(P)}$), the natural logarithm of the GP amplitude ($\ln{(amp)}$),  two decay timescales for the variability (quality factors, $\ln{(Q1)}$ and $\ln{(Q2)}$), and a mixture term (Mix). All parameters evolved under uniform priors with only physical limitations.

We show the best-fit parameters with uncertainties in Extended Table \ref{tab:bestfitParams}. We also show the phase-folded light curve with the best-fit and sample-fit model transits and representative sectors with the GP fit in Figure \ref{fig:GPmodel} and representative transits in Extended Figure \ref{fig:individuals}.

To test for consistency, we fit individual transits of \planetname, locking $P$, $T_0$, and limb-darkening parameters, but allowing $R_P/R_*$ and $a/R_*$ to vary alongside a Matern GP kernel to handle the stellar variability. Overall, 94\% of the $R_P/R_*$ results and 100\% of the $a/R_*$ are within 2$\sigma$ of the global fit. We then repeated this process locking $R_P/R_*$ and $a/R_*$ but letting the individual transit time vary. As with the depths and durations, 88\% of times were within 2$\sigma$ of expectations. 

We fit the ground-based transits using the {\tt BATMAN} model as above, but replaced the GP with a linear model to handle out-of-transit variations and locked other parameters to the fit of the \tess\ data excluding $R_P/R_*$ and $T_0$. The resulting transit fits are shown in Figure \ref{fig:LCOtransits}, and the inferred transit parameters are shown in Extended Table \ref{tab:transitDepths}. The estimated transit depth was consistent across all observed wavelengths, and all agreed with the {\it TESS} depth to $< 1\sigma$.

\subsection{RV analysis:}

We fit the relative RVs from HPF. Fitting just the HPF data allowed us to minimize the effects of long-term variations in the stellar signal. We fit only three parameters, the planet's semi-amplitude ($K$), a velocity offset ($\gamma$), and the jitter ($\sigma_J$). All parameters evolved under uniform priors with only physical limits. The jitter term is added in quadrature with the measurement uncertainties. Other parameters were locked to the results from the transit and we assumed zero eccentricity. Using $K$, we fit for the mass and find an upper limit on the mass of \planetname\ of $0.3M_J$ (95\%). The fitted mass posterior agrees with the distribution of planets of similar radii (see Extended Figure \ref{fig:massRadius}), however, the best-fit mass suggests \planetname\ is low density compared to other large planets with mass measurements, and models predict young super-Earths and mini-Neptunes should be similar in radius to \planetname \cite{Rogers2023}. 

Adding in the APOGEE or IGRINS relative RVs did not significantly impact the mass limit, but increased the jitter term. This is as expected given that the data spans a time where we expect much larger changes in the stellar signal and there were too few points to model any zero-point offsets between instruments. Loosening the eccentricity constraints or allowing for more complex linear or sinusoidal jitter terms increases the mass limit, but under any assumptions the planet is $<2M_J$ (95\%).

\subsection{False positive analysis:}

We consider each false-positive scenario individually below. 

{\it Transit signal originates from instrumental artifacts:} The transit is recovered in both \tess\ and ground-based data across different sites, filters, and instruments.

{\it Transit signal originates from stellar variability:} While the stellar variability amplitude is large (1-10\%; Figure~\ref{fig:GPmodel}), the signal is spread over $\simeq$11.3 days; the stellar signal is weaker than the planet signal over the 6-hour transit duration. The stellar variability profile changes radically between \tess\ sectors (Figure~\ref{fig:GPmodel}), yet the transit signal remains constant. Finally, the transit is recovered from ground-based data at multiple wavelengths, while stellar signals will depend on wavelength.

{\it Transit signal is from a background or foreground star:}
   \starname\ has moved $\sim$305\,mas over the 14 years spanned by the high-resolution data. This is more than the inner working angle of the data ($\sim$100\,mas), so we can use archival imaging to search for any stars that were behind or in front of the target at the time of our observations. We show the locations of \starname\ at each of the imaging epochs in Extended Figure~\ref{fig:patientImaging}. This rules out any background/foreground source with $\Delta K< 5$ mag.

   We can use the transit depth and shape to set limits on the brightness of sources that can reproduce the transit by considering the depth and duration of the largest possible companion. From this, the ratio of the ingress or egress duration to the duration from first to third contact (start of ingress to start of egress) limits the radius ratio of the transit source regardless of any diluting flux \cite{Seager_MO2003, Vanderburg2019}. This calculation suggests if the transit signal comes from another star, it must be $\Delta T \leq 1.7$ mag fainter ($T\le14.5$ mag) at 99\% confidence. This is far brighter than the limits set by the imaging.

{\it \planetname\ is an eclipsing binary or brown dwarf:} A basic fit to the RVs rules out anything with mass $>0.3M_J$. Even under pessimistic assumptions about the eccentricity and stellar variability, all brown dwarf masses ($>13M_J$) are easily rejected. Further, the Sonora Bobcat models \cite{Marley2021} predict an object with a brown-dwarf mass ($\sim$13 $M_J$) at this age would be $\sim2.0\,R_J$, inconsistent with our fitted radius ($\simeq1R_J$).

{\it Transit signal is from an eclipsing binary bound to \starname:} To set additional limits on bound companions, we combine the high resolution images, RVs, and \gaia\ imaging using \texttt{MOLUSC} (Multi-Observational Limits on Unseen Stellar Companions)\cite{Wood2021}. \texttt{MOLUSC} generates synthetic companions and compares them to the observational data. Since we are only interested in companions that could reproduce the transit, we reject any companion that would not land within the photometric aperture used to recover the ground-based transits (2.7\arcsec). This also rules out the bound companion (4\arcsec) and any nearby sources detected in \gaia. \texttt{MOLUSC} generated $1,000,000$ sample stars and $\sim 98.4\%$ were rejected. 

The remaining companions consistent with the observational data generally have small mass ratios ($q <0.06$) with long orbital periods ($\log_{10}(P/days) = 6.31^{+1.46}_{-1.91}$). Applying the brightness constraints above removes all but 800 companions (0.08\%). The remaining binaries are all $\Delta J<1.5$; such systems would appear as a second set of lines in our IGRINS or HPF spectra (SNR$>$100 per epoch) and as an elevated position on the color-magnitude diagram, neither of which were detected.

We can set additional limits using the multi-wavelength ground-based follow-up. If the transit signal were from another star, the depth would vary between bands due to changes in the relative flux contribution with wavelength\cite{Desert2015, THYMEV}. Taking the 95th percentile of depths, we set limits on color combinations from $g$ to $z$. The strongest are in $g-T$, which limit the true host to, at most, 0.67 mag redder than the target. As with the depth and duration constraints, allowed companions would all be detected in our suite of other follow up. 

{\it The transit signal is a persistent flux dip or transient dipper:}
Many disk-bearing young stellar objects show `dips' in their light curves, commonly attributed to dust or gas near the co-rotation radius \cite{Ansdell2016a}. The overwhelming majority of these objects exhibit light curve morphology and quasi-periodicity inconsistent with a planet on a Keplerian orbit. However, PTFO 8-8695 exhibits light curve dips that were originally attributed to a planet \cite{2012ApJ...755...42V}, before it was found to be a transient dipper \cite{Bouma2020}. At least five other systems exhibit similar behavior \cite{2017AJ....153..152S}.

We can reject this scenario by comparing the light curve behavior. PTFO 8-8695 and its brethren are not truly periodic, showing both long-term variations in the dip timing and many events not happening at all \cite{Koen2015}. Depths and durations from dippers (including PTFO 8-8695) vary by a factor of 2-3 over 1-2 month windows \cite{Ciardi2015, Yu2015}. None of these systems show a flat bottom and sharp ingress/egress, the hallmarks of non-grazing planetary transits. Transits of \planetname\ are consistent over the full four years of data; all transits show consistent depths and durations, and events occur at the expected times. The ingress/egress and bottom of \planetname\ transits all follow the expected shape of a planetary system with a low impact parameter.

\begin{table}
\centering
\label{tab:bestfitParams}
\begin{tabular}{lc} 
\hline 
\hline 
\multicolumn{2}{c}{Measured Parameters} \\ 
\hline 
$T_0$ (BJD) & $2458821.8257 \pm 0.0028$ \\
$P$ (days) & $8.834976 \pm 2.4\times10^{-5}$ \\
$R_P/R_{\star}$ & $0.0673^{+0.0024}_{-0.0023}$ \\
$b$ & $0.26^{+0.22}_{-0.18}$ \\
$\rho_{\star}$ ($\rho_{\odot}$) & $0.284^{+0.031}_{-0.068}$ \\
$q_{1,1}$ & $0.29^{+0.16}_{-0.13}$ \\
$q_{2,1}$ & $0.329^{+0.096}_{-0.1}$ \\
\hline 
\multicolumn{2}{c}{GP Parameters} \\ 
\hline 
$\ln{P}$ & $2.42605^{+0.00052}_{-0.0005}$ \\
$\ln{amp}$ & $-6.87^{+0.37}_{-0.13}$ \\
$\ln{Q1}$ & $178.4 \pm 120.0$ \\
$\ln{Q2}$ & $0.0128^{+0.0211}_{-0.0096}$ \\
Mix & $4.5^{+3.7}_{-3.9}$ \\ 
\hline 
\multicolumn{2}{c}{Derived Parameters} \\ 
\hline 
$a/R_{\star}$ & $11.81^{+0.42}_{-1.0}$ \\
$i$ ($^{\circ}$) & $88.76^{+0.87}_{-1.0}$ \\
$T_{14}$ (days) & $0.2465^{+0.005}_{-0.0041}$ \\
$T_{23}$ (days) & $0.2119^{+0.0041}_{-0.005}$ \\
$R_P$ ($R_J$) & $0.97 \pm 0.057$ \\
$a$ (AU) & $0.0813^{+0.0048}_{-0.0081}$ \\
$T_{\mathrm{eq}}$ (K) & $805.0^{+38.0}_{-21.0}$ \\
\hline 
\end{tabular} 
\caption{Parameters of \planetname} 
\end{table}

\begin{table}
    \centering
    \begin{tabular}{lccr}
    \hline 
    \hline
    Parameter & Host & Companion & Source \\
    \hline 
    \multicolumn{4}{c}{Identifiers} \\
    \hline
    IRAS & 04125+2902 & ... & IRAS \\
    Gaia & 164800235906366976 & 164800235906367232 & \Gaia\ DR3 \\
    TIC & 56658270 & 56658273 & \tess\ \\
    TOI & 6963 & ... & \tess\ \\
    2MASS & J04154278+2909597 & J04154269+2909558 & 2MASS \\
    ALLWISE & J041542.77+290959.5 & ... & ALLWISE\\
    \hline
    \multicolumn{4}{c}{Astrometry} \\ 
    \hline
    $\alpha$ & $63.928341$ & $63.927944$ & \Gaia\ DR3 \\
    $\delta$ & $29.166539$ & $29.165482$ & \Gaia\ DR3 \\
    $\mu_{\alpha}$ (mas yr$^{-1}$ ) & $12.104 \pm 0.035$ & $12.56 \pm 0.16 $ & \Gaia\ DR3 \\
    $\mu_{\delta}$ (mas yr$^{-1}$ ) & $-18.145 \pm 0.023$ & $-17.19 \pm 0.11$ & \Gaia\ DR3 \\
    $\pi$ (mas) & $6.247 \pm 0.027$ & $6.26 \pm 0.12 $ & \Gaia\ DR3 \\
    \hline
    \multicolumn{4}{c}{Photometry} \\
    \hline
    $TESS$ (mag) & $12.8131 \pm 0.0068$ & $15.619 \pm 0.013$ & \tess\ \\
    $G_{Gaia}$ (mag) & $13.9814 \pm 0.0017$ & $17.4450 \pm 0.0015$ & \Gaia\ DR3 \\
    $BP_{Gaia}$ (mag) & $15.595 \pm 0.007 $ & $20.64 \pm 0.13$ & \Gaia\ DR3\\
    $RP_{Gaia}$ (mag) & $12.738 \pm 0.004 $ & $15.832 \pm 0.007 $ & \Gaia\ DR3\\
    $u'$ (mag) & $19.718 \pm 0.003$ & $23.382 \pm 0.609$& SDSS DR16\\
    $g'$ (mag) & $16.811 \pm 0.004$ & $24.661 \pm 0.889$& SDSS DR16\\
    $r'$ (mag) & $14.652 \pm 0.003$ & $19.397\pm 0.03$& SDSS DR16\\
    $i'$ (mag) & $13.648 \pm 0.001$ & $16.736 \pm 0.006$& SDSS DR16\\
    $z'$ (mag) & $12.568 \pm 0.003$ & $15.205 \pm 0.006$& SDSS DR16\\
    $J$ (mag) & $10.709	\pm 0.029$ & $12.726 \pm 0.034$ & 2MASS\\
    $H$ (mag) & $9.756	\pm 0.030$ & $11.807 \pm 0.041$ & 2MASS\\
    $K_s$ (mag) & $9.376 \pm 0.028$ & $11.288 \pm 0.028$ & 2MASS\\
    $W1$ (mag) & $9.192 \pm	0.030$ & ... & ALLWISE \\
    $W2$ (mag) & $9.096	\pm 0.032$ & ... & ALLWISE \\
    $W3$ (mag) & $8.714	\pm 0.037$ & ... & ALLWISE \\
    $W3$ (mag) & $4.739	\pm 0.026$ & ... & ALLWISE \\
    \hline
    \multicolumn{4}{c}{Physical Properties} \\
    \hline
    $P_{rot}$ (days) & $11.31 \pm 0.06$ & ... & This work \\ 
    $v \sin i_\star$(km s$^{-1}$) & $7.1 \pm 0.5$ & ... & This work\\
    $i_\star$($^\circ$) & $>78$ & ... & This work \\
    $F_{bol}$ (erg cm$^{-2}$ s$^{-1}$ ) & $(5.8 \pm 0.6) \times 10^{-10}$ & ... & This work\\
    $T_{eff}$ (K) & $3912 \pm 72$ & ... & This work\\
    $A_V$ (mag) & $2.71\pm0.20$ & & This work \\
    $R_\star$ (\rsun ) & $1.48 \pm 0.07$ & ... & This work\\
    $L_\star$ (\lsun)  & $0.466 \pm 0.041$ & ... & This work\\
    $M_\star$ (\msun)  & $0.70 \pm 0.04$ & $0.17 \pm 0.04$ & This work\\
    Age (Myr) & \multicolumn{2}{c}{$3.3^{+0.6}_{-0.5}$}  &  This work \\
    \hline
    \end{tabular}
    \caption{Properties of the host star \starname\ and wide binary companion \companionname}
    \label{tab:stellarParameters}
\end{table}

\begin{figure}
    \centering
    \includegraphics[width=0.99\textwidth]{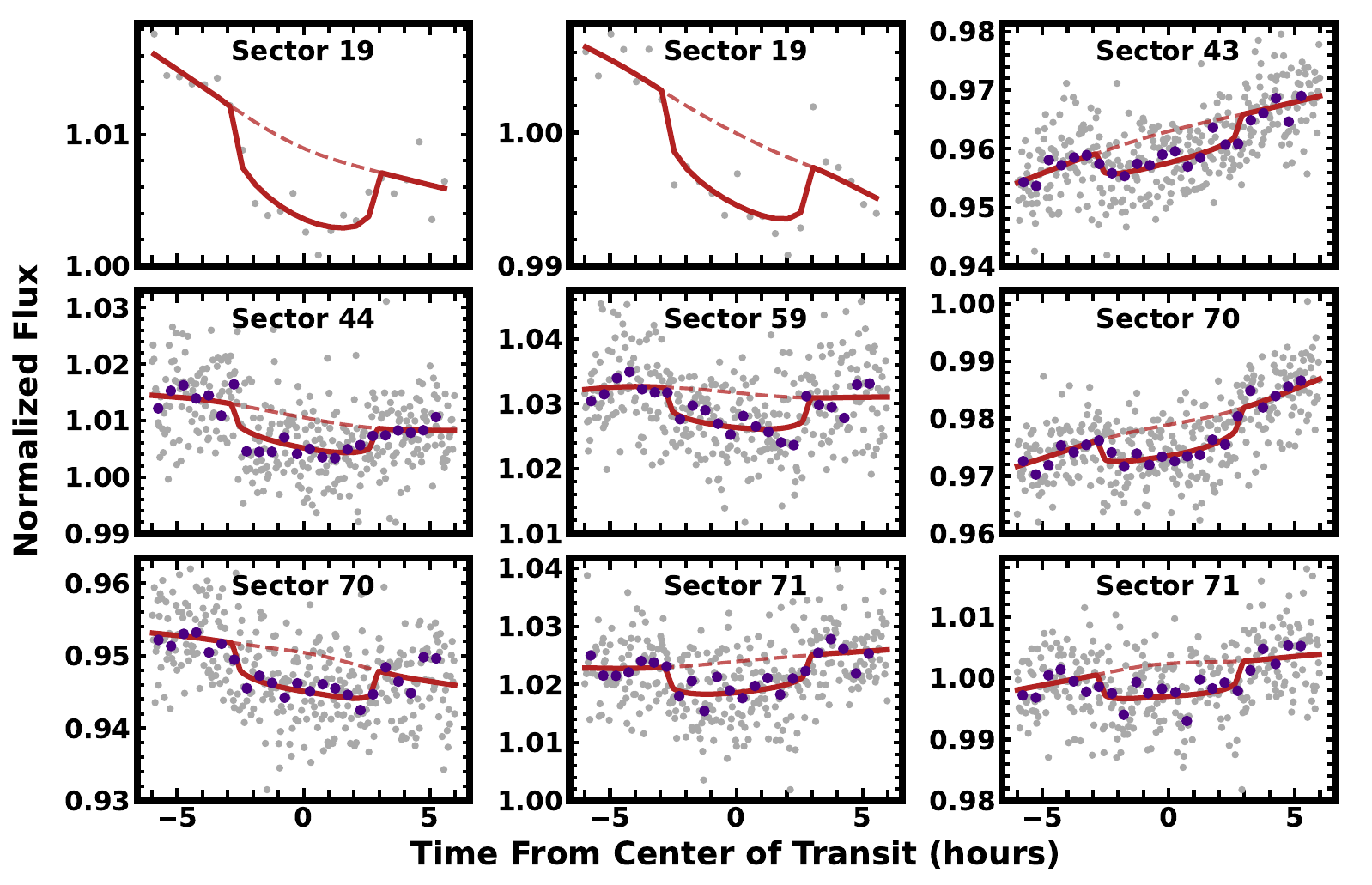}
    \caption{Representative transits from each \tess\ sector (gray points). SPOC data in sectors 43, 44, 59, 70, and 71 are binned to 30 minute intervals (purple points) for a fair comparison to sector 19 (30-minute FFI only). The combined GP and transit model is shown as the solid red line with a dashed line showing the GP fit without the transit.}
    \label{fig:individuals}
\end{figure}

\begin{table}
    \centering
    \caption{Transit Depths at Different Wavelengths} 
    \begin{tabular}{ccccc} 
    \hline 
    \hline 
    Filter & $R_p/R_*$ & $T_{0,O}-T_{0,C}$ (days) & a & b \\ 
    \hline 
    \vspace{-0.3 cm}\\
SDSS $g$ & $0.0647^{+0.0088}_{-0.009}$ & $0.002^{+0.005}_{-0.005}$ & $-0.0014^{+0.0011}_{-0.0011}$ & $0.0018^{+0.0116}_{-0.011}$\\
SDSS $i$ & $0.0696^{+0.006}_{-0.0063}$ & $0.004^{+0.003}_{-0.003}$ & $-0.0027^{+0.0007}_{-0.0007}$ & $0.033^{+0.0062}_{-0.0061}$\\
PS $z$ & $0.0703^{+0.0019}_{-0.0019}$ & $0.0^{+0.002}_{-0.002}$ & $-0.0019^{+0.0004}_{-0.0004}$ & $0.0085^{+0.0017}_{-0.0017}$\\
    \hline 
    \end{tabular}
    \label{tab:transitDepths}
\end{table}

\begin{figure}
    \centering
    \includegraphics[width=0.98\textwidth]{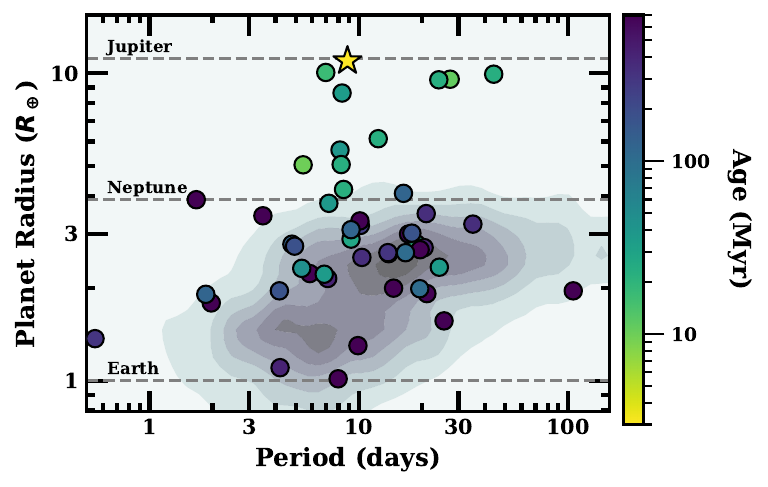}
    \caption{The contour distribution of (primarily older) planets discovered by \Kepler\ and {\it K2} \cite{planetContours} overlapped with the young ($<700$ Myr) transiting systems discovered in a cluster or association. The young planets are colored by their approximate age in log space. \planetname\ is marked with a star and sits high in radius compared to older systems. Planet properties from \cite{EXOFOP_Kepler}. }
    \label{fig:youngplanetDist}
\end{figure}

\begin{figure}
    \centering
    \includegraphics[width=0.98\textwidth]{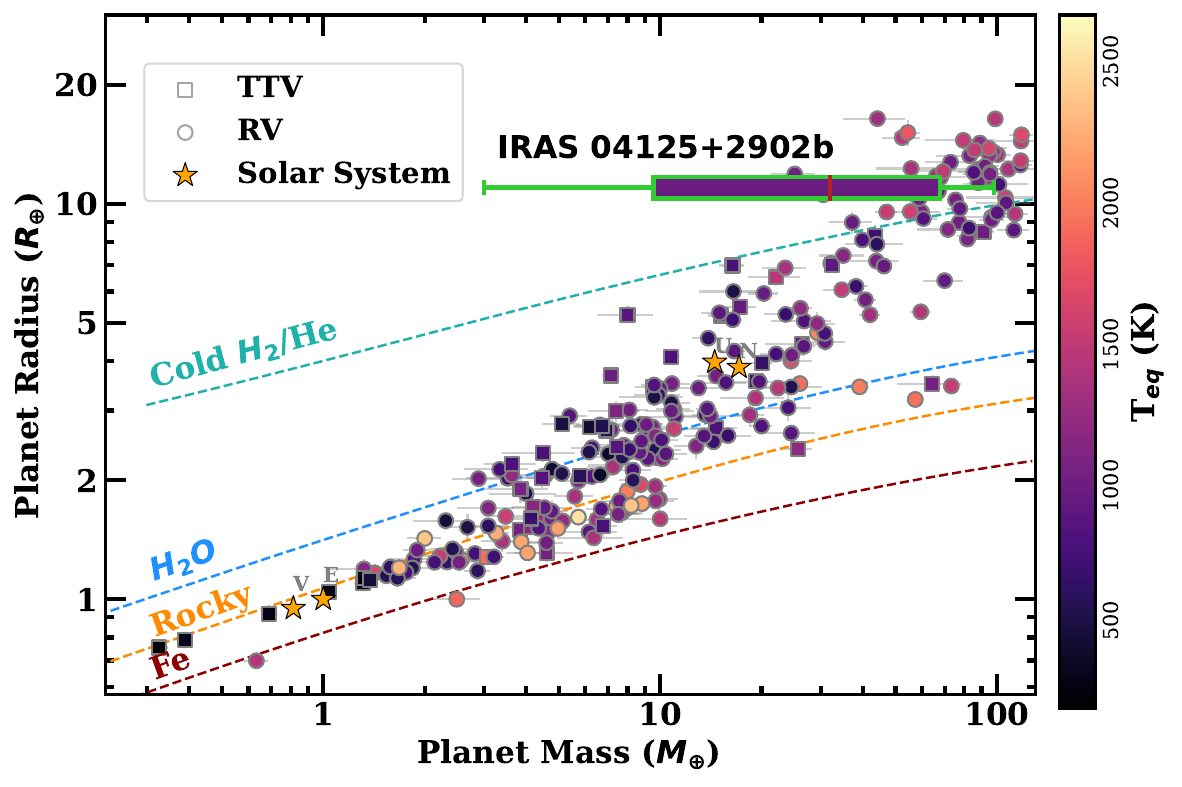}
    \caption{Mass and radius distribution of planets with orbital periods $<20$ days and relative uncertainties of $<25\%$ for mass and $<8\%$ for radius. Squares represent TTV masses, and circles represent RV masses. Points are colored based on their equilibrium temperature. Venus, Earth, Uranus, and Neptune are marked as yellow stars for reference. Dashed lines indicate different planet compositions; iron (red), rocky (orange), water (blue), and cold H2/He (turquoise) (\cite{Zeng2019}). \planetname\ is represented by the box plot. The red line indicates the median mass measurement, with the box and whiskers equating to the $1\sigma$ and $2\sigma$ uncertainties, respectively. The height of the box is equivalent to the $1\sigma$ radius uncertainty. While the mass of \planetname\ does agree with planets of similar radii, \planetname\ is likely a low density planet, and a more precise mass is needed to determine how it compares in parameter space.}
    \label{fig:massRadius}
\end{figure}

\begin{figure}
    \centering
    \includegraphics[width=0.50\textwidth]{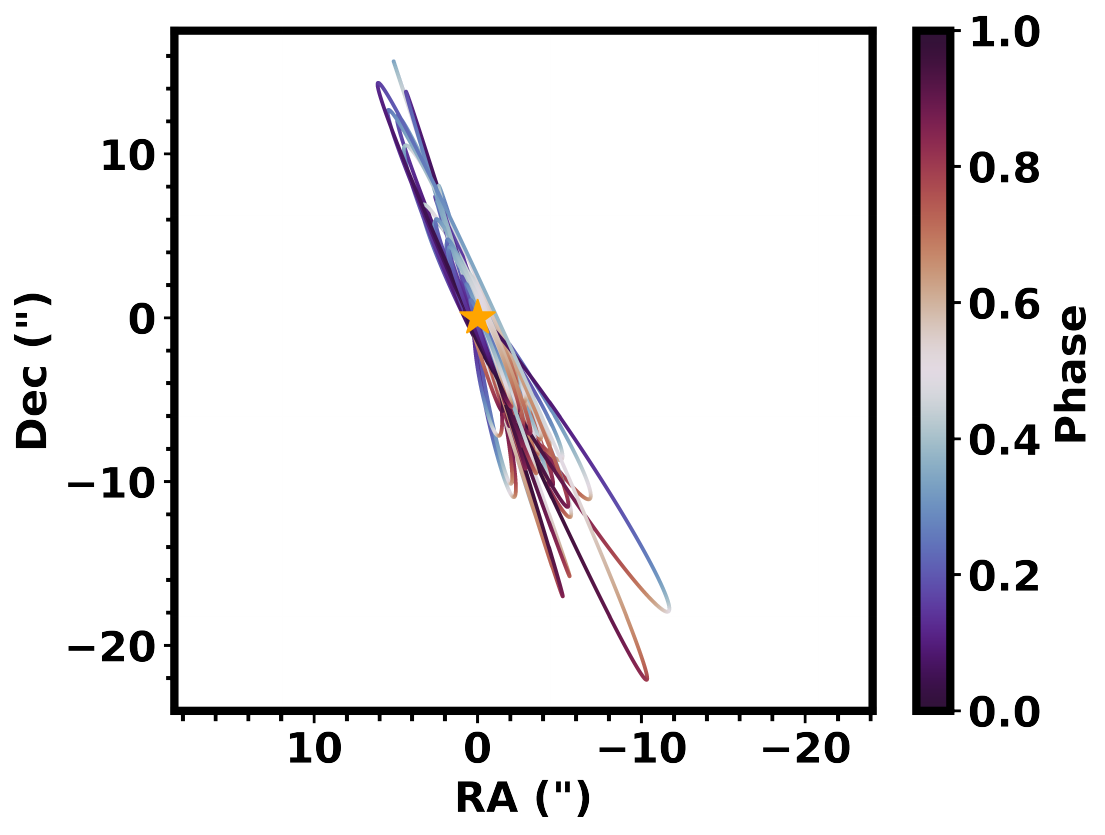}
    \includegraphics[width=0.41\textwidth]{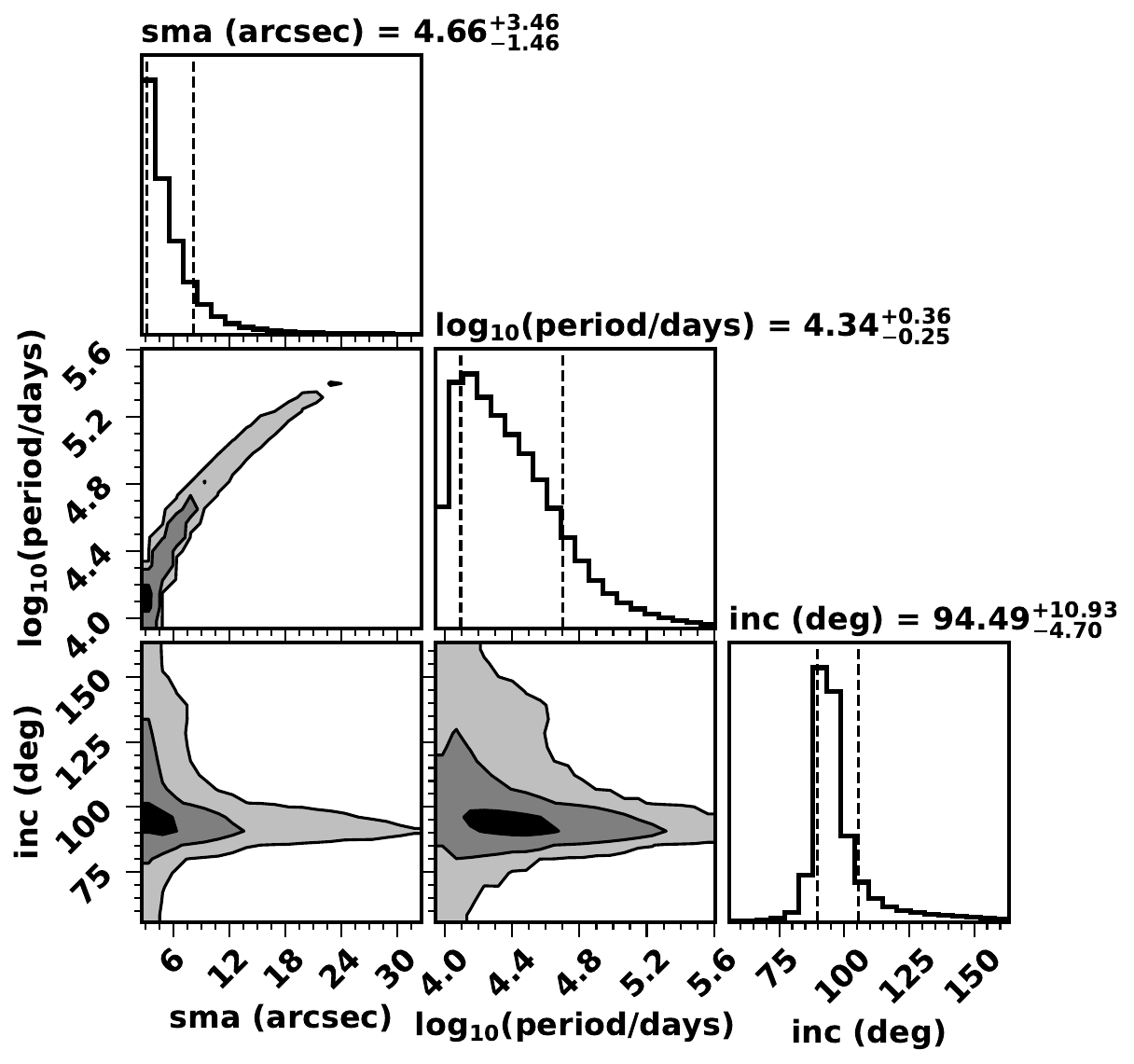}
    \caption{Left: 100 random fits of the orbit of \companionname\ colored by the orbital phase. The orange star represents the location of the host star, \starname. Right: Corner plot of the parameters of the wide binary companion. The dashed lines represent the 16th and 84th percentile of the distribution. 99\% of points are shown for clarity. The inclination of the orbit is highly consistent with edge-on.}
    \label{fig:companionInclination}
\end{figure}

\begin{figure}
    \centering
    \includegraphics[width=0.98\textwidth]{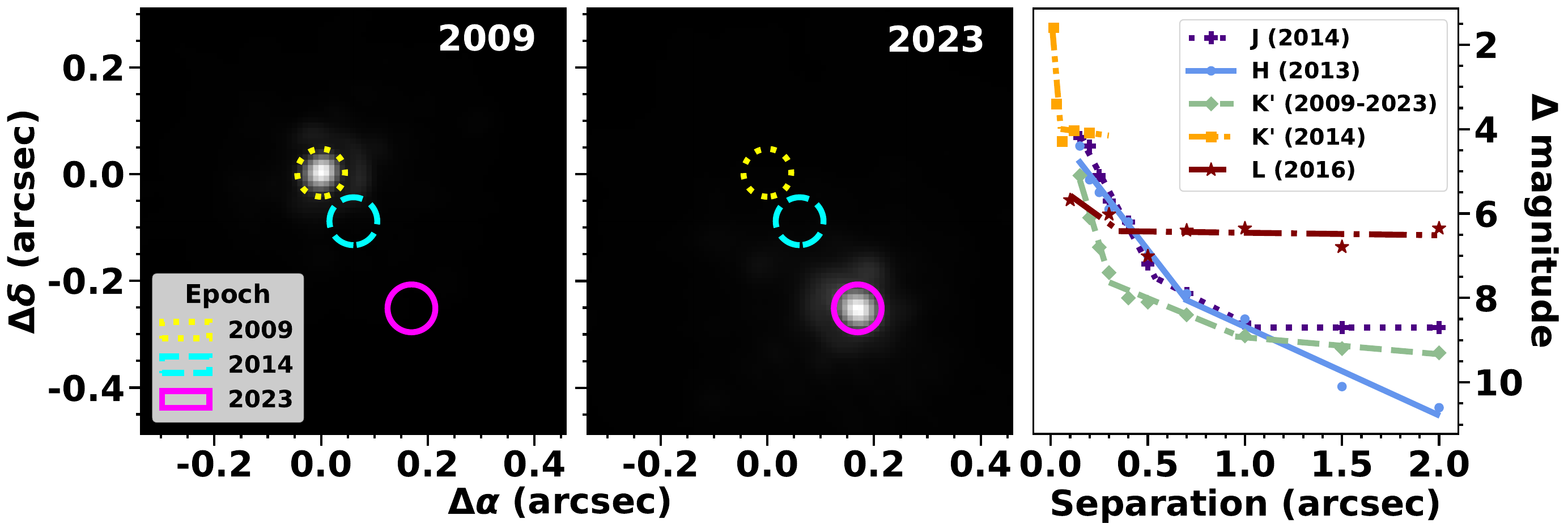}
    \caption{AO images from Keck/NIRC2 of \starname\ across 2 epochs with circles representing the previous/future locations: 2009 (dotted yellow), 2014 (dashed blue), and 2023 (solid pink). Over the 14 year period, \starname\ moves $\sim305$ mas (6 $\lambda$$/D$), which is greater than the PSF width of the star, and we can see there are no stars coming in or out of view that could be causing the observed transit signal. Right: Contrasts from high-resolution data. We show the deepest data for each band, but additionally separating $K$-band imaging (green) and $K$-band NRM (yellow) as they cover different angular separations. A simple break-point linear fit for each band is shown for clarity.}
    \label{fig:patientImaging}
\end{figure}

\end{methods}

\bibliography{bibs/extrabib.bib, bibs/taurusPaper.bib, bibs/andrewbib.bib} 
\bibliographystyle{naturemagfixed}

\begin{addendum}

\item[Acknowledgments] The authors thank Gregory Herczeg for providing the SNIFS spectrum of the host star. We also thank Sarah Blunt for her comments on the manuscript.
M.G.B. was supported by NSF Graduate Research Fellowship (DGE-2040435), the NC Space Grant Graduate Research Fellowship, and the TESS Guest Investigator Cycle 5 program (21-TESS21-0016, PI D.D.). A.W.M. was supported by the NSF CAREER program (AST-2143763). M.D.F. is supported by an NSF Astronomy and Astrophysics Postdoctoral Fellowship (AST-2303911). D. D. acknowledges support from the TESS Guest Investigator Program (80NSSC23K0769). M.F. was supported by NASA’s exoplanet research program (XRP 80NSSC21K0393). 
Funding for the TESS mission is provided by NASA’s Science Mission Directorate. We acknowledge the use of public TESS data from pipelines at the TESS Science Office and at the TESS Science Processing Operations Center. Resources supporting this work were provided by the NASA High-End Computing (HEC) Program through the NASA Advanced Supercomputing (NAS) Division at Ames Research Center for the production of the SPOC data products. TESS data presented in this paper were obtained from the Mikulski Archive for Space Telescopes (MAST) at the Space Telescope Science Institute. This work makes use of observations from the LCOGT network. This paper uses observations from the Habitable Zone Planet Finder on the Hobby Eberly Telescope at the McDonald Observatory.

\item[Author Contributions] M.G.B. identified the planet candidate, preformed the transit and false positive analysis, helped organize further analysis and observations, made most figures for the paper, and wrote large portions of the manuscript.
A.W.M. estimated parameters of the host star and companion, organized LCO follow-up, helped with analysis of the ground-based photometry, and wrote large portions of the manuscript.
A.V. wrote the light curve extraction pipeline to help identify the target and consulted on the pipeline's outputs.
D.K. analyzed the radial velocity from HPF and wrote the portion of the manuscript describing the analysis.
A.K. helped organize the HPF observations, contributed to the LCO observation time, contributed NIRC2 observation time, analyzed the AO images, and contributed to the binary analysis.
M.A. analyzed the disk and wrote the portion of the manuscript describing the analysis. 
L.P. analyzed the binary's orbit and wrote the portion of the manuscript describing the analysis.
G.M. fit the IGRINS data and extracted the radial velocities.
G.M., D.J., and E.S. provided the IGRINS data. 
S.A. and C.E. provided the SMA data.
A.B. calculated the rotation period and wrote the portion of the manuscript that describes the analysis. 
K.C., R.P.S., F.M., E.P. and C.N.W. contributed LCO observation time and reduced the data.
M.D.F. aided in the Keck/NIRC2 observations.
D.D. provided funding for the project during the planet discovery and characterization. 
A.F. created the schematic representation of the system.
M.F. analyzed the \vsini.
A.I.L.M. analyzed the TTVs.
B.M.T. contributed to the LCO observation time, contributed NIRC2 observation time.
P.C.T. contributed code to make the planet demographics plots. 
J.J. leads the SPOC which provided the calibrated FFI data and PDC and SAP light curves used in the analysis.
J.J., D.L., G.R., S.S., R.V., and J.N.W. are TESS mission architects.
D.C., Z.E., D.R.R, A.S., J.D.T., and J.N.V. are TESS mission contributors.

\item[Data Availability] We provide all reduced light curves and spectra as supplementary data products to the manuscript. 

\item[Competing Interests] The authors declare that they have no competing financial interests. 
 
\item[Correspondence] Correspondence and requests for materials should be addressed to Madyson Barber.
 
\item[Code availability] We provide access to a GitHub repository including all code created for the analysis of this project that is not already publicly available.

\end{addendum}

\end{document}